\documentclass{aa}
\usepackage{natbib}
\usepackage{txfonts}
\usepackage{graphicx}
\usepackage[dvips]{rotating}
\bibpunct{(}{)}{;}{a}{}{,}

\def\ion{\mathrm{ion}}
\def\uv{\mathrm{uv}}

\def\exc{\mathrm{exc}}
\def\abs{\mathrm{abs}}
\def\bol{\mathrm{bol}}

\def\cm{\hbox{cm}}

\def\AU{\hbox{AU}}

\def\eff{\mathrm{eff}}

\begin{document}
\title{ISO spectroscopy of disks around Herbig Ae/Be stars 
 \thanks{Based on observations with ISO, an ESA project with instruments 
         funded by ESA Member States (especially the PI countries: France, 
         Germany, the Netherlands and the United Kingdom) and with the 
         participation of ISAS and NASA.}} 
\titlerunning{Disks Around HAEBE Stars}
\authorrunning{Acke \& v.d.~Ancker}
\author{Bram~Acke\inst{1} \and Mario~E.~van~den~Ancker\inst{2}}
\institute{Instituut voor Sterrenkunde, KULeuven, Celestijnenlaan 200B, 
3001 Leuven, Belgium\\ 
\email{Bram.Acke@ster.kuleuven.ac.be}
\and
European Southern Observatory, Karl-Schwarzschild Strasse 2, D-85748
Garching bei M\"unchen, Germany\\ 
\email{mvandena@eso.org}}
\date{DRAFT, \today}

\abstract{We have investigated the infrared spectra of all 46 Herbig Ae/Be 
stars for which spectroscopic data are available in the ISO data archive. 
Our quantitative 
analysis of these spectra focuses on the emission bands at 3.3,
6.2, ``7.7'', 8.6 and 11.2 micron, linked to polycyclic aromatic
hydrocarbons (PAHs), the nanodiamond-related features at 3.4 and 3.5 
micron, the amorphous 10 micron silicate band and the crystalline
silicate band at 11.3 micron. We have detected PAH emission in 57\%
of the Herbig stars in our sample. Although for most of these sources
the PAH spectra are similar, there are clear examples of differences in 
the PAH spectra within our sample which can be explained by differences 
in PAH size, chemistry and/or ionization. Amorphous silicate emission 
was detected in the spectra of 52\% of the sample stars, amorphous 
silicate absorption in 13\%. We have detected crystalline silicate
emission in 11 stars (24\% of our sample), of which four (9\%) also
display strong PAH emission.
We have classified the sample sources according to the strength of their 
mid-IR energy distribution.  
The systems with stronger mid-infared (20--100~$\mu$m) excesses relative 
to their near-infrared (1--5~$\mu$m) excess display significantly more 
PAH emission than those with weaker mid-infrared excesses. 
There are no pronounced differences in the behaviour of the silicate feature 
between the two groups. 
This provides strong observational support for the disk models by 
\citet{dullemond01}, in which systems with a flaring disk geometry 
display a strong mid-infrared excess, whereas those with disks that 
are strongly shadowed by the puffed-up inner rim of the disk only 
display modest amounts of mid-infrared emission.  Since the silicates 
are expected to be produced mainly in the warm inner disk regions, no 
large differences in silicate behaviour are expected between the 
two groups. In contrast to this, the PAH emission is expected to 
be produced mainly in the part of the disk atmosphere that is directly 
exposed to radiation from the central star. In this model, self-shadowed 
disks should display weaker PAH emission than flared disks, consistent 
with our observations.
}

\maketitle


\section{Introduction}

Although the optical---sub-mm energy distribution of Herbig Ae/Be (HAEBE) stars
has been well explored by previous authors \citep[e.g.][]{hillenbrand},
the chemical and mineralogical composition of the dust remained
poorly studied until the 1995 launch of the {\it Infrared Space
Observatory} \citep[ISO,][]{kessler}. This first possibility to
study the complete infrared spectrum of these objects in detail revealed
a large variety in dust properties, from small aromatic hydrocarbons
to silicate dust.  Moreover, some sources were shown to contain
partially crystalline dust grains, similar to those found in
comets in our own solar system \citep{waelkens96,malfait,
  malfait99,vandenancker00a, vandenancker00b, meeus01}.

ISO played a major role in opening up the field of infrared spectroscopy. 
The emission features at 3.3, 6.2, 7.7, 8.6 and 11.2 micron, found in
ISO spectra of many HAEBE stars and
previously known as the unidentified infrared (UIR) bands, are 
generally attributed to polycyclic aromatic hydrocarbons
\citep[PAHs,][]{leger, allamandola}. PAHs
are a large family of molecules, for which the fundamental ingredients
are polycyclic benzene rings. PAH molecules are thought to be excited
by far-ultraviolet (UV) photons. The absorption of such a photon induces a
transition of the PAH molecule to an upper electronic state. The
excited molecule then makes rapid transitions to a lower electronic
state, leaving most of the initially absorbed energy in the form of
vibrational energy in the CC and CH bonds. PAHs cool down by infrared
(IR) emission in bands linked to these vibrational modes \citep[][and
  references therein]{peeters}. The 3.3 $\mu$m band is due to CH
bond stretching vibrations, the 6.2 $\mu$m band is linked to the CC
stretching mode and the 7.7 $\mu$m band corresponds to a combination
of the CC stretching and the CH in-plane bending modes. The 8.6 and
11.2 $\mu$m features are linked to the CH in-plane and out-of-plane
bending modes respectively \citep[][and references therein]{vermeij}. 

In a few sources, emission bands at 3.4 and 3.5 micron are
observed. These spectral features are attributed to the CH stretching
modes of hydrogenated nanodiamonds
\citep[NANs,][]{guillois,vankerckhoven}.

The infrared spectra of HAEBEs also contain silicate features.
The broad feature at 10 micron has been attributed to the SiO
bonds in warm, small amorphous silicate grains like olivine. 
The feature can appear in emission or in absorption in HAEBE spectra. 
The spectral signature at 11.3 micron, as well as many features at
longer wavelengths which are not included in the analysis presented in
this paper, is linked 
to crystalline silicates \citep{malfait, malfait99, vandenancker00a,
vandenancker00b, bouwman00, bouwman01}.

The spectral energy distribution (SED) of HAEBE stars is characterized
by the presence of an IR flux excess, due to thermal emission of
circumstellar matter. The geometry of the circumstellar dust has been
the subject of a long-lasting debate \citep[e.g.][]{hillenbrand,
  berrilli, hartmann, bohm94, grinin94, grinin96, grady, corcoran98,
  mannings97}. For late-B, A and F stars, the evidence for the
presence of disk-like 
geometries is generally accepted. For early-B stars, the matter is
less clear. In these systems, the dissipation time scale of the
circumstellar spherical envelope is of the order of the
pre-main-sequence life time. Disks as well as spherical
envelopes might be present.

In their study of the ISO spectra of 14 isolated Herbig Ae/Be stars, 
\citet[][henceforth M01]{meeus01} classified their sample 
into two groups, based on the shape of the SED. 
\textit{Group I} contains the sources in which a rising mid-IR
(20--100$\mu$m) flux excess is observed; these sources have an SED that can
be fitted with a power-law and a black-body continuum. \textit{Group
II} sources have a more modest mid-IR excess; their SEDs can be
reconstructed by a power-law only. M01 suggest phenomenologically that
this classification represents different geometries of the
circumstellar disk: group I sources have flared disks, group II
members have flat disks.

\citet[][henceforth D02]{dullemond02} and \citet[][henceforth
  DD04]{dullemond04} have modelled young stellar disks 
with a self-consistent model based on 2-D radiative transfer coupled
to the equation of vertical hydrostatics. The model consists of a disk
with an inner hole ($\sim$0.5 AU), a puffed-up inner rim and an outer
part. The outer part of the disk can be flared \citep[as
  in][]{chiang}, but can also lie entirely in the shade of the inner 
rim. The SEDs of flared disks display a strong mid-IR flux excess,
while self-shadowed disks have a much more modest mid-IR excess. D02
explains quantitatively the 
difference in SED shape in HAEBEs (as expressed by the classification
of M01) as the result of a different disk
geometry; group I sources have flared disks, group II sources have
flat self-shadowed disks.

In this paper we investigate a possible link between the shape of the
SED ---a proxy for the geometrical distribution of the circumstellar
matter--- and the strength and profiles of circumstellar infrared
emission bands in HAEBE systems.
All spectra of HAEBE stars taken by ISO are investigated,
amongst others spectra that have never been published before. It is therefore
the most complete sample of near-IR spectra of HAEBE stars ever
investigated as a whole. 
Our study is not only qualitative, but contains also, as opposed to
many other articles, a quantitative analysis of the infrared spectra of
HAEBEs.


\section{The data set}

\subsection{The sample}

Our list of HAEBEs is based on the catalogue of \citet[][henceforth
  T94]{the}. We restricted ourselves to Table 1 and 2 of that 
article. We enlarged the sample with HAEBEs studied by
  \citet[][henceforth M98]{malfait}, that satisfy the criteria postulated by T94. 
The resulting list was cross-correlated with the ISO data archive; we selected
all objects for which ISO spectra were available with an ISO pointing
within a distance of $\pm$5 arcseconds of the position given by T94
and M98. 

We eliminated some of the remaining sources from the sample; TY CrA
was omitted because the infrared emission emanates from the ``TY
CrA bar'' \citep{siebenmorgen}, which is not directly associated
to the star. The object also does not display H$\alpha$ emission, which
indicates that this star is no longer actively accreting.
From a comparison between the ISO-SWS spectrum of MWC 137 and
photometric measurements from the MSX Point Source Catalog
\citep{egan}, it appears that there is an offset of 12\arcsec\, between the 
position indicated in T94 and the position of the infrared source. 
Only a small portion of the object's flux is seen in
the SWS spectrum. We therefore discarded this spectrum from our
analysis. 
A similar problem was encountered for the SWS spectrum of IRAS 12496-7650. A
mispointing of the ISO instrument of about 17\arcsec\, compared to the
2MASS point-source-catalogue position \citep{cutri} caused an 
anomaly between the ISO SWS-spectra and the IR photometry. However, 
the PHT-S spectra for this source were not affected due to the larger
aperture of the ISO-PHOT instrument (see Sect.~\ref{datred}). 
51 Oph and MWC 300 were also removed from the list; these are probably
evolved objects and not HAEBEs \citep[][respectively]{vandenancker01,
  molster}. 

The resulting sample of 46 sources is tabulated in Table~\ref{sample}.  
The columns contain respectively the object's name, the type of
ISO spectrum (see Sect.~\ref{datred}), the ISO Observation Sequence
Number (OSN), the right ascension (RA), the declination (Dec)
(both epoch 2000 coordinates), the date and starting time (UT start)
of the measurement and the total integration time T in seconds. The
objects are sorted by increasing RA. The given coordinates are the
pointing coordinates of ISO for that source.

\begin{table*}
\caption{ The sample of HAEBEs used in this study, based on
  \citet{the} and \citet{malfait}. For each object, the observed ISO
spectra are given; S01 refers to SWS AOT01, P40 to PHT-S and C04 to
CAM04 spectra. \label{sample}}
\begin{center}
\begin{tabular}{cccccccc}
 \hline
 \multicolumn{8}{c}{\bf Sample stars}\\ \hline \hline
 \multicolumn{1}{c}{Object} &
 \multicolumn{1}{c}{AOT} &
 \multicolumn{1}{c}{OSN} &
 \multicolumn{1}{c}{RA (2000)} &  
 \multicolumn{1}{c}{Dec (2000)} &
 \multicolumn{1}{c}{Date} &
 \multicolumn{1}{c}{UT start} &
 \multicolumn{1}{c}{T}  \\ 
  &   &  & $h\ \ m\ \ s\ $ & $^{\circ}\ \ \ '\ \ \ "$  &  & $h\ \ m\ \ s\ $  & [s] \\
\hline
\object{V376 Cas}      & S01 & 43501514 & 00 11 26.6 & $+$58 50 04 & 24-Jan-1997& 17:00:14 & 3554     \\    
\object{VX Cas}        & P40 & 58704023 & 00 31 30.5 & $+$61 58 51 & 25-Jun-1997& 15:42:53 &  364     \\    
\object{Elias 3-1}     & S01 & 67301306 & 04 18 40.7 & $+$28 19 16 & 19-Sep-1997& 01:03:47 & 3454     \\    
\object{AB Aur}        & S01 & 68001206 & 04 55 45.7 & $+$30 33 06 & 26-Sep-1997& 05:26:35 & 3454     \\    
\object{HD 31648}      & S01 & 83501201 & 04 58 46.1 & $+$29 50 38 & 27-Feb-1998& 19:47:48 & 3454     \\    
\object{UX Ori}        & P40 & 85801453 & 05 04 30.0 & $-$03 47 14 & 22-Mar-1998& 19:37:29 &  364     \\    
\object{HD 34282}      & S01 & 83301240 & 05 16 00.5 & $-$09 48 34 & 25-Feb-1998& 18:57:46 & 1912     \\    
\object{HD 34700}      & S01 & 66302638 & 05 19 41.4 & $+$05 38 43 & 09-Sep-1997& 11:10:28 & 1912     \\    
                       & P40 & 63602294 & 05 19 41.4 & $+$05 38 42 & 13-Aug-1997& 12:51:00 &  172     \\    
\object{HD 35187}      & S01 & 69501139 & 05 24 01.2 & $+$24 57 36 & 10-Oct-1997& 23:44:54 & 1912     \\    
\object{BF Ori}        & P40 & 70101958 & 05 37 13.3 & $-$06 35 01 & 17-Oct-1997& 08:33:28 &  364     \\    
\object{RR Tau}        & P40 & 86603163 & 05 39 30.5 & $+$26 22 26 & 30-Mar-1998& 21:38:08 &  364     \\    
\object{Z CMa}         & S01 & 72201607 & 07 03 43.2 & $-$11 33 07 & 07-Nov-1997& 04:42:28 & 3454     \\    
\object{HD 95881}      & S01 & 10400818 & 11 01 57.8 & $-$71 30 52 & 29-Feb-1996& 05:04:40 & 3462     \\    
                       & P40 & 10400919 & 11 01 57.8 & $-$71 30 52 & 29-Feb-1996& 06:03:04 &  140     \\    
\object{HD 97048}      & S01 & 14101343 & 11 08 04.6 & $-$77 39 17 & 06-Apr-1996& 10:15:08 & 3462     \\    
                       & S01 & 61801318 & 11 08 04.6 & $-$77 39 17 & 26-Jul-1997& 08:30:31 & 6538     \\    
                       & P40 & 07900309 & 11 08 04.6 & $-$77 39 17 & 04-Feb-1996& 06:06:00 &  172     \\    
                       & P40 & 14101580 & 11 08 04.6 & $-$77 39 17 & 06-Apr-1996& 11:26:28 &  364     \\    
                       & P40 & 62501510 & 11 08 04.6 & $-$77 39 17 & 02-Aug-1997& 04:36:23 &  620     \\    
                       & C04 & 71801836 & 11 08 03.5 & $-$77 39 17 & 03-Nov-1997& 02:54:09 & 2208     \\    
                       & C04 & 71901688 & 11 08 04.6 & $-$77 39 17 & 04-Nov-1997& 04:17:02 & 2846     \\    
\object{HD 100453}     & P40 & 26000131 & 11 33 05.6 & $-$54 19 29 & 02-Aug-1996& 17:38:16 &  140     \\    
                       & S01 & 26000230 & 11 33 05.7 & $-$54 19 29 & 02-Aug-1996& 17:41:20 & 1912     \\    
\object{HD 100546}     & S01 & 07200660 & 11 33 25.5 & $-$70 11 42 & 28-Jan-1996& 13:35:29 & 1044      \\  
                       & S01 & 27601036 & 11 33 25.3 & $-$70 11 42 & 07-Jul-1996& 11:37:28 & 1912     \\    
                       & P40 & 10400537 & 11 33 25.7 & $-$70 11 42 & 29-Feb-1996& 04:14:02 &  140     \\    
\object{HD 104237}     & S01 & 10400424 & 12 00 06.0 & $-$78 11 34 & 29-Feb-1996& 03:40:48 & 1834     \\    
                       & S01 & 23300524 & 12 00 05.1 & $-$78 11 34 & 07-Jul-1996& 11:37:28 & 1912     \\    
                       & P40 & 23300625 & 12 00 05.1 & $-$78 11 34 & 07-Jul-1996& 12:10:04 &  140     \\    
                       & P40 & 53300118 & 12 00 05.1 & $-$78 11 34 & 02-May-1997& 02:32:49 &  364     \\    
\object{IRAS12496-7650}& S01 & 23300112 & 12 53 15.9 & $-$77 07 02 & 06-Jul-1996& 19:27:51 & 3455     \\    
                       & P40 & 07901717 & 12 53 16.1 & $-$77 07 02 & 04-Feb-1996& 12:56:06 &  140     \\    
\object{HD 135344}     & S01 & 10401575 & 15 15 48.4 & $-$37 09 16 & 29-Feb-1996& 10:10:12 & 1834     \\    
                       & P40 & 10401742 & 15 15 48.4 & $-$37 09 16 & 29-Feb-1996& 11:47:12 &  140     \\    
                       & P40 & 10401876 & 15 15 48.4 & $-$37 09 16 & 29-Feb-1996& 12:11:18 &  140     \\    
\object{HD 139614}     & S01 & 29701542 & 15 40 46.3 & $-$42 29 53 & 09-Sep-1996& 03:28:45 & 1912     \\    
                       & P40 & 10402322 & 15 40 46.5 & $-$42 29 55 & 29-Feb-1996& 14:45:38 &  140     \\    
\object{HD 141569}     & S01 & 62802937 & 15 49 57.6 & $-$03 55 16 & 05-Aug-1997& 04:40:44 & 1912     \\    
                       & P40 & 62701662 & 15 49 57.7 & $-$03 55 17 & 04-Aug-1997& 08:02:39 &  236     \\    
\object{HD 142527}     & S01 & 10402046 & 15 56 42.1 & $-$42 19 24 & 29-Feb-1996& 12:42:16 & 1834     \\    
                       & P40 & 10402547 & 15 56 42.1 & $-$42 19 24 & 29-Feb-1996& 15:42:20 &  140     \\    
\object{HD 142666}     & S01 & 10402952 & 15 56 40.1 & $-$22 01 41 & 29-Feb-1996& 18:23:12 & 3462     \\    
                       & S01 & 44901283 & 15 56 40.1 & $-$22 01 39 & 07-Feb-1997& 19:12:46 & 6538     \\    
                       & P40 & 10402847 & 15 56 40.1 & $-$22 01 41 & 29-Feb-1996& 17:59:06 &  140     \\    
\object{HD 144432}     & S01 & 45000284 & 16 06 58.0 & $-$27 43 08 & 08-Feb-1997& 08:28:30 & 6539     \\    
                       & P40 & 10402662 & 16 06 58.0 & $-$27 43 10 & 29-Feb-1996& 16:01:26 &  140     \\    
\object{HR 5999}       & S01 & 28901506 & 16 08 34.3 & $-$39 06 19 & 01-Sep-1996& 02:46:48 & 1912     \\    
                       & P40 & 28901748 & 16 08 34.2 & $-$39 06 18 & 01-Sep-1996& 04:28:10 &  140     \\    
                       & C04 & 45800955 & 16 08 34.3 & $-$39 06 19 & 16-Feb-1997& 17:01:44 & 2198     \\    
\object{Wra 15-1484}   & S01 & 29901001 & 16 27 14.9 & $-$48 39 27 & 10-Sep-1996& 20:04:23 & 1912     \\    
\object{HD 150193}     & S01 & 08200444 & 16 40 17.9 & $-$23 53 45 & 07-Feb-1996& 07:02:19 & 1044     \\    
                       & P40 & 64102335 & 16 40 17.9 & $-$23 53 44 & 18-Aug-1997& 13:13:35 &  236     \\    
\object{AK Sco}        & S01 & 28902101 & 16 54 44.8 & $-$36 53 18 & 01-Sep-1996& 07:24:34 & 1140     \\    
                       & P40 & 64402829 & 16 54 44.8 & $-$36 53 17 & 21-Aug-1997& 13:00:15 &  236     \\    
 \hline
\end{tabular}
\end{center}
\end{table*}

\begin{table*}
\begin{center}
\begin{tabular}{cccccccc}
 \hline
 \multicolumn{8}{c}{\bf Sample stars (continued)}\\ \hline \hline
 \multicolumn{1}{c}{Object} &
 \multicolumn{1}{c}{AOT} &
 \multicolumn{1}{c}{OSN} &
 \multicolumn{1}{c}{RA (2000)} &  
 \multicolumn{1}{c}{Dec (2000)} &
 \multicolumn{1}{c}{Date} &
 \multicolumn{1}{c}{UT start} &
 \multicolumn{1}{c}{T}  \\ 
  &   &  & $h\ \ m\ \ s\ $ & $^{\circ}\ \ \ '\ \ \ "$  &  & $h\ \ m\ \ s\ $  & [s] \\
\hline 
\object{CD$-$42$^\circ$11721}   & S01 & 08402527 & 16 59 06.8 & $-$42 42 08 & 09-Feb-1996& 16:26:56 & 1816     \\    
                        & S01 & 64701904 & 16 59 05.8 & $-$42 42 15 & 24-Aug-1997& 11:11:16 & 1912     \\    
                        & P40 & 28900460 & 16 59 06.8 & $-$42 42 08 & 31-Aug-1996& 17:58:54 &  364     \\    
\object{HD 163296}      & S01 & 32901191 & 17 56 21.4 & $-$21 57 20 & 10-Oct-1996& 21:05:00 & 3454     \\    
                        & P40 & 32901192 & 17 56 21.3 & $-$21 57 20 & 10-Oct-1996& 22:03:18 &  140     \\    
\object{HD 169142}      & S01 & 13601359 & 18 24 30.0 & $-$29 46 50 & 01-Apr-1996& 09:46:06 & 1834    \\    
                        & P40 & 13601437 & 18 24 29.8 & $-$29 46 50 & 01-Apr-1996& 10:30:40 &  140    \\    
\object{MWC 297}        & S01 & 70800234 & 18 27 39.5 & $-$03 49 52 & 23-Oct-1997& 18:56:23 & 1912    \\    
\object{VV Ser}         & P40 & 47800913 & 18 28 47.9 & $+$00 08 40 & 08-Mar-1997& 16:27:17 &  364    \\    
\object{R CrA}          & S01 & 14100458 & 19 01 53.9 & $-$36 57 10 & 06-Apr-1996& 04:48:02 & 1834    \\    
                        & S01 & 70400558 & 19 01 53.9 & $-$36 57 10 & 19-Oct-1997& 20:12:45 & 1912    \\    
                        & P40 & 11501230 & 19 01 53.4 & $-$36 57 04 & 11-Mar-1996& 11:04:57 &  140    \\    
\object{T CrA}          & S01 & 33402096 & 19 01 58.8 & $-$36 57 49 & 16-Oct-1996& 04:15:14 & 3454    \\    
                        & S01 & 68900196 & 19 01 58.8 & $-$36 57 49 & 04-Oct-1997& 18:11:15 & 3454    \\    
                        & P40 & 14100562 & 19 01 58.8 & $-$36 57 49 & 06-Apr-1996& 06:07:04 &  140    \\    
\object{HD 179218}      & S01 & 32301321 & 19 11 11.2 & $+$15 47 17 & 05-Oct-1996& 03:27:54 & 3454    \\    
                        & C04 & 36401031 & 19 11 11.3 & $+$15 47 16 & 14-Nov-1996& 16:11:58 & 2240    \\    
\object{WW Vul}         & S01 & 17600305 & 19 25 59.0 & $+$21 12 30 & 10-May-1996& 23:54:57 & 1834    \\    
                        & P40 & 17600465 & 19 25 59.0 & $+$21 12 31 & 11-May-1996& 01:33:17 &  140    \\    
                        & P40 & 51300108 & 19 25 58.6 & $+$21 12 31 & 12-Apr-1997& 03:50:24 &  364    \\    
\object{BD+40$^\circ$4124}     & S01 & 35500693 & 20 20 28.3 & $+$41 21 51 & 05-Nov-1996& 18:19:26 & 3454    \\    
                         & P40 & 15900568 & 20 20 28.3 & $+$41 21 51 & 24-Apr-1996& 03:39:08 &  140    \\    
\object{LkH$\alpha$ 224}& S01 & 85800502 & 20 20 29.2 & $+$41 21 27 & 22-Mar-1998& 09:49:31 & 1912    \\    
\object{LkH$\alpha$ 225}& S01 & 85800403 & 20 20 30.4 & $+$41 21 27 & 22-Mar-1998& 08:51:15 & 3454    \\    
\object{PV Cep}         & S01 & 14302273 & 20 45 54.0 & $+$67 57 36 & 08-Apr-1996& 13:37:16 & 1834    \\    
                         & P40 & 14302274 & 20 45 54.0 & $+$67 57 36 & 08-Apr-1996& 14:08:34 &  140    \\    
\object{HD 200775}      & S01 & 33901897 & 21 01 36.8 & $+$68 09 49 & 21-Oct-1996& 03:36:10 & 3454    \\    
                         & C04 & 10702605 & 21 01 37.1 & $+$68 09 53 & 03-Mar-1996& 12:41:22 & 3874    \\    
\object{V645 Cyg}       & S01 & 26301850 & 21 39 58.2 & $+$50 14 22 & 06-Aug-1996& 07:29:22 & 1912    \\    
\object{BD+46$^\circ$3471}     & C04 & 54101787 & 21 52 34.1 & $+$47 13 44 & 10-May-1997& 17:20:17 & 2804    \\    
\object{SV Cep}         & S01 & 28800703 & 22 21 33.0 & $+$73 40 24 & 30-Aug-1996& 18:44:21 & 3454    \\    
                         & P40 & 56201203 & 22 21 33.1 & $+$73 40 27 & 31-May-1997& 09:49:36 &  364    \\    
\object{MWC 1080}       & S01 & 26301659 & 23 17 25.8 & $+$60 40 43 & 06-Aug-1996& 04:43:40 & 1912    \\    
                         & S01 & 28301459 & 23 17 25.8 & $+$60 40 43 & 25-Aug-1996& 23:15:57 & 1912    \\  
 \hline
\end{tabular}
\end{center}
\end{table*}

\subsection{Data Reduction \label{datred}}

We retrieved all spectra of HAEBE stars present in the ISO data
archive\footnote{http://www.iso.vilspa.esa.es/ida/index.html}.
Three types of ISO data were used in this study: ISO-SWS
\citep{degraauw}, ISO-PHT \citep{lemke} and ISO-CAM \citep{cesarsky}.

ISO-SWS was the Short Wavelength Spectrometer aboard ISO (2.38--45.2
$\mu$m). This instrument consisted of 4 sets of 12 detectors. Each set
of detectors covered a different wavelength region ([2.38:4.08],
[4.08:12.0], [12.0:29.0] and [29.0:45.2] $\mu$m respectively). Not
only the material of which the 
detectors were made differed from band to band, also the aperture
sizes were larger at longer wavelengths (from 14\arcsec $\times$
20\arcsec\, around 3 $\mu$m to 20\arcsec $\times$ 33\arcsec\, around 30
$\mu$m). The data of different detectors and wavelength regions are
calibrated independently.

The source's spectrum was scanned in time, in a way that all 12
detectors of one set scanned through the same wavelength region
twice. We only used the SWS data in which the whole spectrum was scanned
(scanning mode AOT01). In the standard pipeline reduction, dark
currents (residual signal, not related to the actual astrophysical
target) are subtracted from the raw data. The detector signal is
corrected by multiplication with a responsitivity curve. Furthermore, a
wavelength and flux calibration are carried out. We used the
resulting product of the latest version of the off-line processing
pipeline reduction (OLP 10). 
After this basic reduction, we manually removed bad data.
This includes the suppression of deviating detectors (detectors which
measure spectra that are too dissimilar with respect to the spectra
observed by the other detectors operating in the 
same wavelength region) and artefacts like glitches (spikes
in intensity, spread out over several data points due to the
detector's memory effects).
The latter are caused by infalling highly energetic interplanetary
particles. A glitch that exceeds the noise level is easily
recognizable, since it is only present in one of the two scan
directions of the detector and its shape is characteristic.

To diminish the scatter in the data, the calibrated flux
levels in a wavelength interval are averaged and all detectors are
scaled to this 
value. The noise distribution is strongly non-Gaussian and
asymmetric, among other reasons because of glitches. Since the data will be rebinned
and averaged, removal of the outliers is needed. Outliers with a
deviation of more than three times the noise in the data were
expelled. In the next step the data are rebinned at a spectral
resolution of 150 for low flux sources ($<10$ Jy at 10~$\mu$m) and 500
for high flux sources ($>10$ Jy at 10~$\mu$m).

Sometimes small differences in flux level are present between the
different wavelength regions in the spectrum.
The final phase of the data reduction is to correct for these
jumps. Since most sources in the sample have low fluxes, jumps
are primarily due to errors in the dark current subtraction. Therefore
offsets were applied to correct for the flux differences between
the wavelength intervals. The shifting was done by adding a constant term 
---which was computed to minimize the flux difference in the
overlap region between two adjacent wavelength regions--- to the flux. 
We took the interval between 3.02 to 3.52 micron as the flux
calibration reference and scaled all parts of the spectrum 
between 2.38 and 12.0 micron to that level. For
the region [12.0:29.0] micron, the interval covering [16.5:19.5] micron
was used. We discarded the wavelength range longer than 29 micron,
since the flux levels in this part of the spectrum were too low to be
reliable for many of the sample stars. 

The wavelength region around 12 $\mu$m is difficult to interpret,
since both the end of the previous and the beginning of the next
wavelength range are very
noisy and therefore unreliable. For a few spectra, we found it useful
to extend the wavelength region of the first interval with respect 
to the standard pipeline reduction: from 12 to 12.5
$\mu$m. This was done for \object{V376 Cas}, \object{HD 31648}, \object{Z
CMa}, \object{HD 95881}, \object{HD 104237}, \object{HD 139614}, \object{HD
142666}, \object{HD 150193}, \object{T CrA}, \object{WW Vul},
\object{LkH$\alpha$ 224}, \object{HD 200775} and \object{SV Cep}. Since we
chose to shift the flux levels in 
[2.38:12.0]$\mu$m, and [12.0:29.0]$\mu$m separately, the overlap
region between these two intervals is independent of this
post-reduction flux level correction. In this way, we avoid changing
the shape of spectrum around this wavelength area too much compared to
the original ---unshifted--- data. Not combining the two intervals in
a direct way also prevents the entry of systematic errors in the shape
and line flux of the broad, amorphous 10 micron silicate feature. We
note that each of the narrower features at 3.3, 3.4, 3.5, 6.2, 7.7, 8.6, 11.2 and
11.3 micron are included within one continuous SWS wavelength
interval. Therefore our measurements of line flux and full width at
half maximum of those features will be independent of the procedure
followed for the alignment of the different wavelength ranges.

For \object{HD 97048}, \object{HD 100546}, \object{HD 104237}, \object{HD
142666}, \object{CD$-$42$^\circ$11721}, \object{R CrA}, \object{T CrA} and
\object{MWC 1080}, two ISO-SWS spectra were available. We combined the 
spectra after the reduction to increase the
signal-to-noise ratio. For \object{HD 97048} we decided to keep only
one of the SWS spectra (OSN 61801318), because the second spectrum was
affected by severe instrumental artefacts. 

The ISO-PHT instrument was a photo-polarimeter, which had a
spectroscopic mode (PHT-S). The spatial resolution of the instrument
was 24\arcsec$\times$24\arcsec. The spectra consist of 127 photometric points,
separated into two wavelength intervals ([2.5:4.9]$\mu$m and
[5.8:11.6]$\mu$m). For faint sources, PHT-S was able to 
obtain spectra with a higher S/N ratio than SWS, albeit at much lower
spectral resolution ($\sim$90). We retrieved the resulting spectra from the
latest version of the off-line processing data reduction (OLP 10). For
\object{HD 97048}, \object{HD 104237}, \object{HD 135344} and \object{WW Vul},
two or three PHT-S spectra were available; we merged the spectra. 

The ISO-CAM instrument also had a spectrophotometric mode
(CAM04). This mode made it possible to take a series of images at variable
central wavelengths (between 5 and 15 $\mu$m). We extracted the
spectrum by applying synthetic photometry to the images with a
circular aperture with a diameter of 15\arcsec\, (comparable with the spatial
resolution of SWS and CAM).
Only 5 CAM04 spectra are included in this analysis. Again we used the
results of the latest version of the off-line processing data
reduction pipeline OLP 10. 

\begin{figure*}
\rotatebox{0}{\resizebox{17cm}{!}{\includegraphics{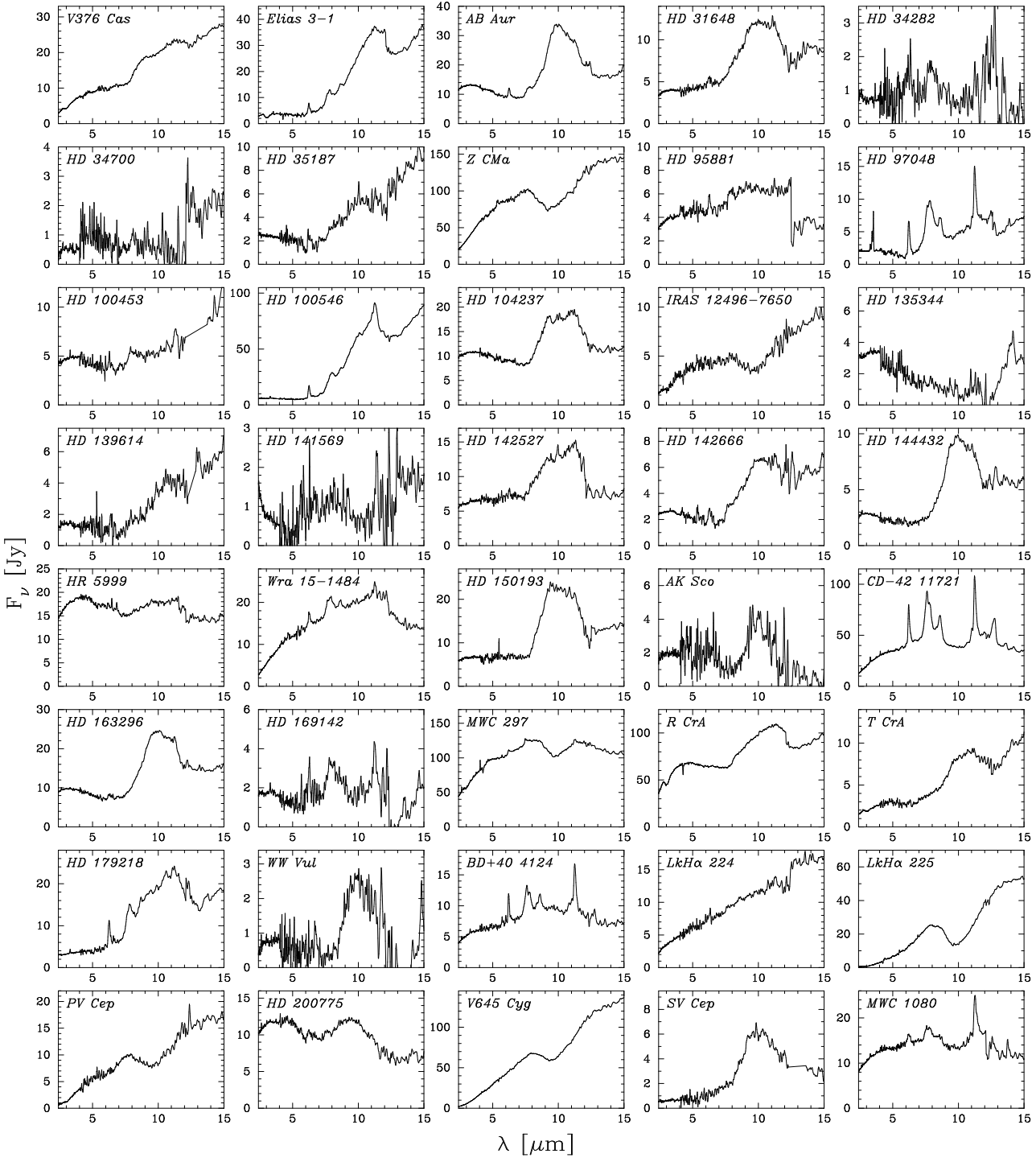}}}
\caption{ The reduced 2.4--15~$\mu$m SWS spectra of the sample stars.}
\label{haebe_sws.ps}
\end{figure*}

\begin{figure*}
\rotatebox{0}{\resizebox{17cm}{!}{\includegraphics{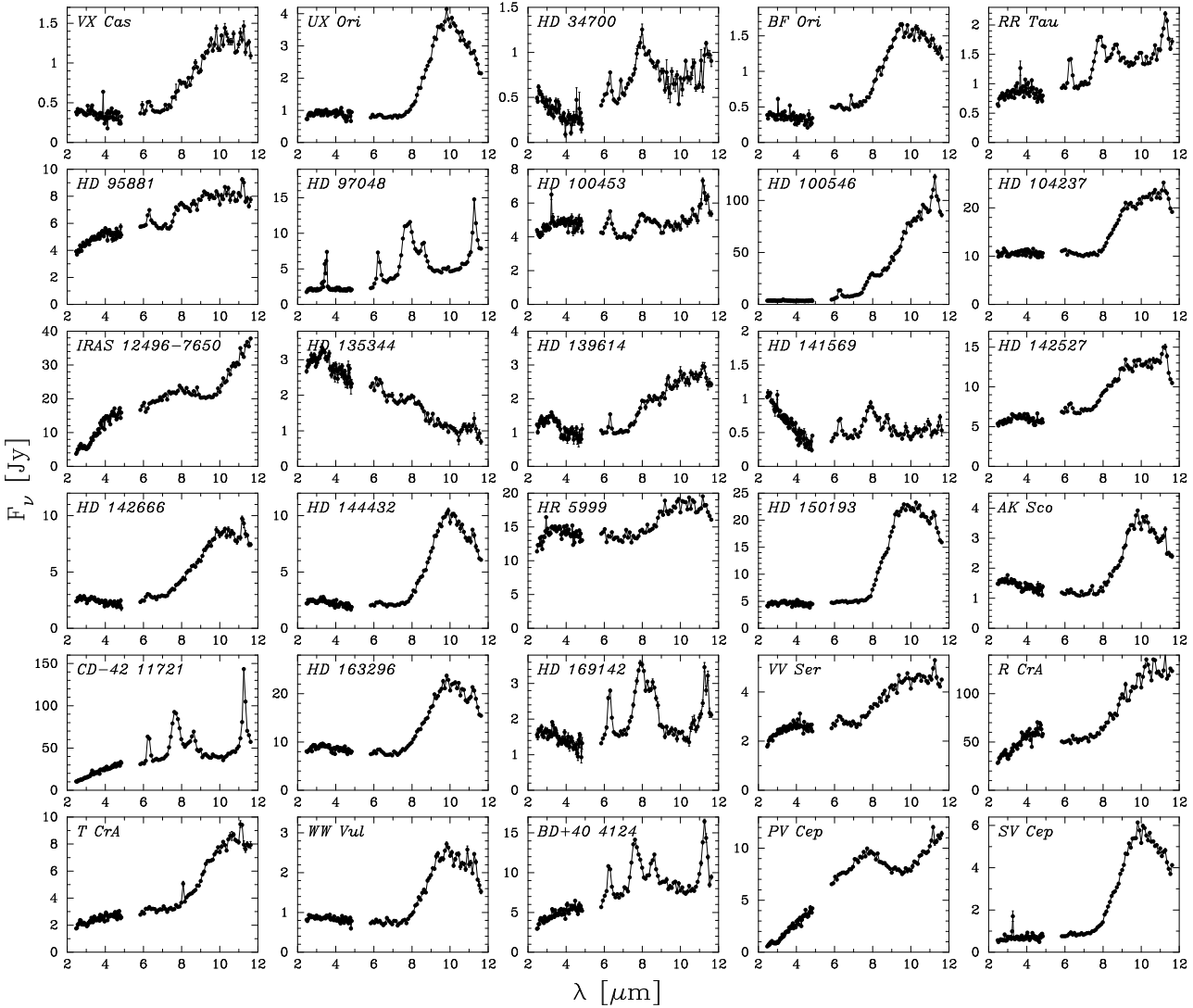}}}
\caption{ The reduced 2-5--11.6~$\mu$m PHT-S spectra of the sample stars.}
\label{haebe_pht.ps}
\end{figure*}

\begin{figure*}
\rotatebox{0}{\resizebox{17cm}{!}{\includegraphics{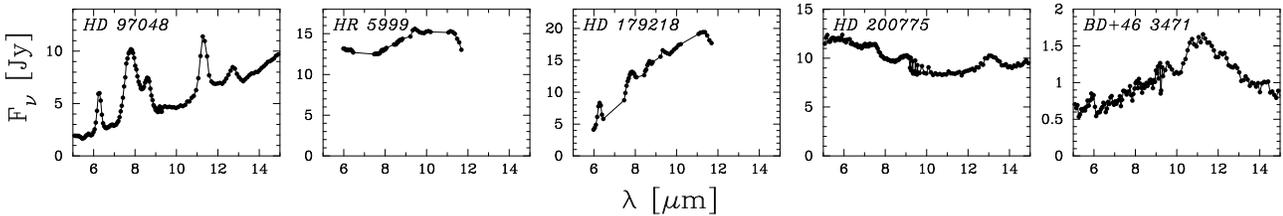}}}
\caption{ The reduced 5--15~$\mu$m CAM spectra of the sample stars.}
\label{haebe_cam.ps}
\end{figure*}

The resulting spectra of the data reduction are compiled in
Fig.~\ref{haebe_sws.ps}, Fig.~\ref{haebe_pht.ps} and
Fig.~\ref{haebe_cam.ps} for ISO-SWS, ISO-PHT and ISO-CAM, 
respectively. 

\subsection{Measurement of the spectral features}

In our analysis, we focused on the infrared emission bands at 3.3,
6.2, 7.7, 8.6 and 11.2 micron, attributed to polycyclic aromatic
hydrocarbons and the features at 3.4 and 3.5 micron, linked to
nanodiamonds. The amorphous silicate feature around 10
micron and the crystalline silicate feature at 11.3 micron were also
included in the analysis. The 11.2 and 11.3 micron bands, when present, are
blended. Due to the low spectral resolution, we cannot
distinguish between the two. Therefore we consider the
blend as one complex, which we call the \textit{11 micron feature}.

To characterize these features in the
spectra, we measured the line flux,  full width at half maximum
(FWHM), peak wavelength, peak flux, continuum flux at the peak
wavelength and the equivalent width (EW).

We wrote an IDL procedure, {\tt measfeat.pro}\footnote{this IDL
  procedure can be downloaded at the following website:
  http://www.ster.kuleuven.ac.be/$\sim$bram/ISO/measfeat.pro and
  http://www.ster.kuleuven.ac.be/$\sim$bram/ISO/measfeat.README}, to
  manipulate the spectra 
in order to determine the desired measurements. The IDL code allows us to 
indicate continuum points by hand. Based on these points a spline function 
is fitted to estimate the continuum. Integrating the continuum subtracted
flux, the line flux and EW of the features were computed.
We repeated the procedure, indicating an 'extreme' continuum, 
to estimate the systematic error introduced by the continuum
determination. It appears that the uncertainty in the line flux varies from about
$\sim$10\% for the PAH features and the 10 micron feature, up to
$\sim$30\% for the 11 micron feature. Furthermore, the IDL program
determined the peak position and peak flux by fitting a Gaussian
function to the continuum-subtracted line profile. 
The peak flux was set equal to the maximum of the fit, the peak
wavelength is the wavelength of this maximum. When the feature was in
absorption, the peak flux was defined as the minimum of the fit. In
the latter case the peak wavelength is the wavelength at which this
minimum occurred. The value of the continuum flux at the peak wavelength 
is also recorded. 
Note that, because of the possible skewness of the features, the peak
wavelength in general does not coincide with the centroid
wavelength. This may lead to systematic differences between the values
used in this paper and values in the literature.

The FWHM was deduced by computing the width at half the peak flux of
the smoothed line profile. The smoothing of the spectrum was done
to diminish the influence of the noise in the spectrum on this
measurement. 

When a feature was not detected, we deduced upper limits for its
line flux and peak flux, and we computed the continuum flux at the
expected central wavelength of the feature. The upper limits were
computed in the following manner; we used the averaged spectrum of
\object{CD$-$42$^\circ$11721} and \object{HD 97048} (both sources with
fairly typical  
PAH spectra; Sect.~\ref{thePAHfeatures}) as a template spectrum for 
the PAH emission, the spectrum of HD 97048 for the 3.4 and 3.5 micron 
NAN emission and
the spectrum of \object{HD 150193} as a template for the 10 micron amorphous
silicate emission. For non-detections, we assumed a peak emission of 5
times the noise on the spectrum at the theoretical central wavelength
of the feature as an upper limit for the peak flux. Scaling the
template spectra to this level, we extracted the upper limits for the
line flux. 

This procedure was applied to the SWS, PHT-S and CAM spectra. Since the
spectral resolution of the PHT-S and CAM data is not even 100, only the
line flux, EW, peak flux and continuum flux at the theoretical central
wavelength were determined for the latter; the other measurements are
not meaningful in this case.

\subsection{Comparison of SWS, PHT-S and ground-based photometry}

\begin{figure}
\rotatebox{0}{\resizebox{3.5in}{!}{\includegraphics{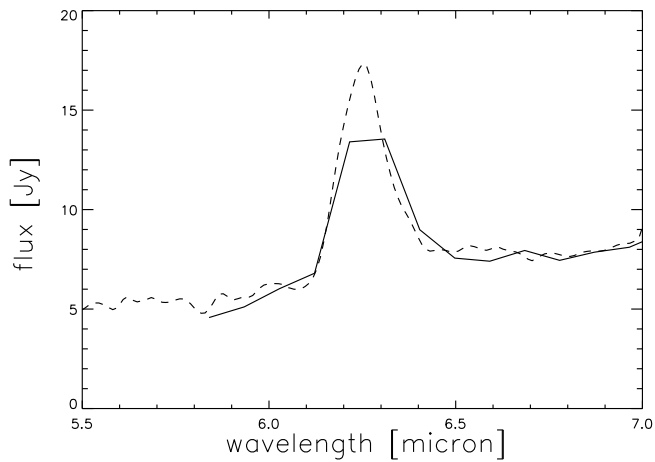}}}
\caption{ The PAH feature at 6.2 micron of \object{HD 100546} in the SWS
spectrum (dashed line) and in the PHT-S spectrum (full line).}
\label{hd100546pah62.ps}
\end{figure}

For 24 sources, SWS as well as PHT-S spectra were available. We
independently measured the desired quantities of the features, when
present in both spectra. Comparing these two sets of values, we
noticed that there is an offset between the two; the computed PHT-S
line fluxes are systematically lower than their SWS counterparts. We
attribute this difference to the fact that at 2.8 times the Nyquist
sample spacing, the PHT-S data are insufficiently sampled in
wavelength to recover the full spectral response. The result of this
is that we will lose flux around sharp gradients in the spectra. As 
an example we show in
Fig.~\ref{hd100546pah62.ps} the PAH 6.2 feature of \object{HD 100546}, 
as it appears in the SWS and PHT-S spectra. The line flux in the PHT-S
spectrum is clearly underestimated. In Table~\ref{LFpht/sws} we
summarize the ratios of the line flux measured in the PHT-S spectrum
over the line flux measured in the SWS spectrum (PHT/SWS) for all
detected features. Note that most of the ratios (81\%) are smaller
than unity, which is consistent with our interpretation. The given
errors are the statistical errors on the measurements. 
Ratios that are larger than one can be due to spatially extended
emission or intrinsically varying PAH emission.

\begin{table*}
\caption{ The ratios of the line flux of the features measured in the
PHT-S spectrum over the line flux measured in the SWS spectrum
LF$_{\mathrm{PHT-S}}$/LF$_{\mathrm{SWS}}$. The given errors are the statistical errors on
the measurements. \label{LFpht/sws}}	
\begin{center}
\begin{tabular}{|c|c|c|c|c|c|}
 \hline
 \multicolumn{6}{|c|}{\bf Comparison PHT-S versus SWS}\\ \hline \hline
 \multicolumn{1}{|c|}{Object} &
 \multicolumn{1}{|c|}{PHT/SWS 3.3} & 
 \multicolumn{1}{|c|}{PHT/SWS 6.2} & 
 \multicolumn{1}{|c|}{PHT/SWS 7.7} &
 \multicolumn{1}{|c|}{PHT/SWS 8.6} &
 \multicolumn{1}{|c|}{PHT/SWS 11} \\ 
\hline 
\object{HD 34700}                  &            &  $<$2.04         &  $<$2.45          &  $<$4.45         &  \\
\object{HD 95881}      		  &            &  0.82 $\pm$     0.06  &  0.48 $\pm$     0.03          & $<$1.33           & $<$ 1.49 \\    
\object{HD 97048}      		  & 0.71 $\pm$     0.01  &  1.06 $\pm$     0.02   &  0.97 $\pm$    0.01  &  0.70 $\pm$     0.01  & 1.23 $\pm$ 0.03  \\    
\object{HD 100453}                 & 0.84 $\pm$      0.12  &  0.59 $\pm$     0.07 &	1.00 $\pm$     0.05          &  $<$0.63          &  \\    
\object{HD 100546}                 & 0.88 $\pm$     0.04  &  0.80 $\pm$     0.01  & 0.89 $\pm$     0.01  &  0.87 $\pm$     0.02 & 2.55 $\pm$      0.10   \\    
\object{HD 135344}     		  &            &  $<$1.75          &            &            &  \\    
\object{HD 139614}                 &            &  $<$2.83          &  $<$2.21          &  $<$1.52         & $<$6.67  \\    
\object{HD 141569}     		  &            &  $<$2.64          &            & $<$4.76           &   \\ 
\object{HD 142527}                 & $>$0.87           &  $<$1.14         &  $<$1.41          &  $<$1.91          & 4.28 $\pm$      0.28  \\    
\object{HD 142666}     		  & $>$0.30           &  3.11 $\pm$      0.24   &  $<$2.00         &            &  0.69 $\pm$      0.10  \\    
\object{HD 144432}                 &            &  $<$1.23          &            &            & $<$ 0.97  \\    
\object{HR 5999}       		  &            &            &      	  &            & 1.55 $\pm$      0.19  \\    
\object{HD 150193}                 &            &            &            &            &  $<$ 1.47  \\    
\object{CD$-$42$^\circ$11721} 		  & 0.65  $\pm$     0.01  &  0.90 $\pm$     0.01 &   0.92 $\pm$    0.01&  0.78 $\pm$    0.01  & 0.75 $\pm$    0.01    \\
\object{HD 163296}                 &            &  0.91 $\pm$     0.08         &	          &            &  1.54 $\pm$      0.10 \\    
\object{HD 169142}                 & $>$1.61           &  0.61 $\pm$     0.05 &  0.95 $\pm$     0.03 &  1.17 $\pm$     0.07  & 1.92 $\pm$      0.23     \\    
\object{BD+40$^\circ$4124}    		  &  $>$1.66           &  0.51 $\pm$     0.01 &  0.59 $\pm$     0.01 &  0.51 $\pm$     0.01 & 1.00 $\pm$     0.03    \\    
 \hline			  
\end{tabular}		  
\end{center}		    
\end{table*}

We calculated the mean ratio of the detected SWS line flux over the
PHT-S line flux for each feature. The values of these
quantities are 1.18, 1.34, 1.17, 1.16, 1.18, 1.08 and 0.58 for PAH
3.3, NAN 3.4, NAN 3.5, PAH 6.2, PAH 7.7, PAH 8.6 and the 11 micron feature
respectively. The PHT-S measurements were adjusted by multiplying with these
mean values. The factors were also applied to correct the upper limits. 

Note that the amorphous 10 micron silicate feature is only partially
sampled in the PHT-S spectra, since the wavelength coverage ends at
11.6 micron. Therefore it is difficult to determine the underlying
continuum in the PHT-S spectra. Nevertheless, we indicated the
continuum in a consistent manner.
Similarly to the PAH features, we computed the line flux ratio of
detected 10 micron emission features in both the SWS and the PHT-S
spectrum, and computed the mean value. By applying this correction
factor (0.86), we compensate for this systematical difference between
the measurements of the SWS and PHT-S 10 micron feature.

The comparison between the CAM and SWS spectra of \object{HD 97048} showed that
the results are equal within a 15\% range. No corrections for the CAM
results were applied. We decided not to use the CAM spectra of
\object{HR 5999}, \object{HD 179218}, \object{HD 200775} and \object{HD 97048}
in the analysis; the first two spectra are inhomogeneously covered in
wavelength, which leads to insufficiently sampled features. The CAM
spectrum of \object{HD 200775} is quite different from the SWS spectrum;
the first displays PAH 
emission, while the latter does not. We relate this to extended
emission, visible in the CAM spectrum because the (synthetic) aperture
is larger than the aperture for the SWS spectrum at these wavelengths.
The CAM spectrum of \object{BD+46$^\circ$3471} is the only spectrum of
this source at our disposal. We added the measurements to our data
set, even though we noticed unusual features in the spectrum, which
could be artefacts (e.g. the ``absorption feature'' at 10 micron).

The ISO spectra were compared to space- and ground-based photometric
measurements from the literature; we plotted the photometric points
from the MSX catalog over the reduced spectra. In general, the mean
deviation between the 
spectra and the photometry is small and non-systematic. Sources where
the anomaly exceeds the 20\% level are \object{HD 104237}, \object{HD
142666}, \object{MWC 297}, \object{BD+40$^\circ$4124}, \object{V645 Cyg} and
\object{MWC 1080}. Except for \object{BD+40$^\circ$4124}, the 
photometric fluxes in these spectra are higher than the SWS or PHT-S
fluxes. We attribute the differences to spatially extended emission and
different beam/aperture sizes, although we cannot exclude the
possibility of intrinsic infrared variability of (some of) our sources.

\subsection{The final numbers \label{finalnumbers}}

After the measurements of the features, we condensed the data set into
a final set of numbers: one measurement for each star and examined
feature. The presence of the features in the HAEBE spectra is
summarized in Table~\ref{grI/IIvsPAH}. The sources are sorted
according to their IR-SED classification (see Sect.~\ref{thesed}).

When only a SWS, PHT-S or CAM spectrum was availaible for a source,
those data were taken. For sources with a SWS and a PHT-S spectrum,
three possibilities arise; the feature is detected in both spectra,
the feature is undetected in both spectra or the feature is detected
in only one of the two spectra. In the first case, the data are merged
using a weighted average, based on the statistical errors on the
measurements. In the second case, the most stringent upper limit for
the feature is taken. In the third case, when the measured line flux
of the detected feature is lower than the upper limit derived from the
other spectrum, the data are consistent and the values of the
measurement are selected. In the inconsistent case (the upper limit is
lower than the measured 
line flux), we set a conservative upper limit for the feature. The
latter only occurred in 4 cases (out of 192 measured
features). This behaviour might be either a result of the different
aperture sizes used by SWS and PHT-S, in which case this could
indicate the existence of spatially extended emission, or a first
indication that the strength of PAH emission may vary with time in
some of our sources. We indicated these measurements in
Table~\ref{grI/IIvsPAH} with $\surd^{ c}$ and $\surd^{ d}$, according
to detection in the PHT-S or SWS spectrum respectively.

The computed line fluxes used in the analysis are listed in
Table~\ref{LF}. Upper limits are indicated as well. Table~\ref{CF}
summarizes the continuum fluxes at the peak wavelength of each
feature.

\begin{table*}[tbh!]
\caption{ The presence of the PAH features at 3.3, 6.2, 7.7 and 8.6
micron, the NAN features at 3.4 and 3.5 micron, the 11 micron
complex (COMP 11) and the amorphous 10 micron silicate feature (Si 9.7) in the
spectra of the sample stars. The sources are classified in 3
groups.
$\surd$: detection; -- : no detection; $n.s.$: no spectrum available
at these wavelengths. $\surd^{ a}$: tentative detection; $\surd^{ b}$: tentative detection in the
PHT-S spectrum; $\surd^{ c}$: feature
undetected in the SWS spectrum; $\surd^{ d}$: feature undetected in
the PHT-S spectrum (see Sect.~\ref{finalnumbers}). For the amorphous silicate feature,
$E$ stands for emission and $A$ for absorption.
\object{BD+40$^\circ$4124}, \object{R CrA} and \object{LkH$\alpha$ 224} have not been classified
based on the diagram in Fig.~\ref{plotroy.ps}, which is indicated by the
question mark.
 \label{grI/IIvsPAH}}	
\begin{center}
\begin{tabular}{|c|c|c|c|c|c|c|c|c|c|}
 \hline
 \multicolumn{10}{|c|}{\bf Group I/II vs IR emission bands.}\\ \hline \hline
 \multicolumn{1}{|c|}{Object} &
 \multicolumn{1}{|c|}{Group} &
 \multicolumn{1}{|c|}{PAH 3.3} & 
 \multicolumn{1}{|c|}{NAN 3.4} & 
 \multicolumn{1}{|c|}{NAN 3.5} &
 \multicolumn{1}{|c|}{PAH 6.2} &
 \multicolumn{1}{|c|}{PAH 7.7} &
 \multicolumn{1}{|c|}{PAH 8.6} &
 \multicolumn{1}{|c|}{COMP 11} &
 \multicolumn{1}{|c|}{Si 9.7} \\ 
\hline \hline 
\object{V376 Cas}                  &  I   &  --     &  --      &  --      &  --     &  --     &   --      &  --      &  $E$  \\
\object{Elias 3-1}     		  &  I   & $\surd$ & $\surd$  & $\surd$  & $\surd$ & $\surd$ &	$\surd$  & $\surd$  &  $E$  \\
\object{AB Aur}        		  &  I   &  --     &  --      &  --      & $\surd$ & $\surd$ &	$\surd$  & $\surd$  &  $E$  \\
\object{HD 34282}      		  &  I   &  --     &  --      &  --      &  --     & $\surd$ &	 --      &  --      &   --  \\
\object{HD 34700}                  &  I   &  --     &  --      &  --      & $\surd$ & $\surd$ &	$\surd$  &  --      &   --  \\
\object{HD 97048}      		  &  I   & $\surd$ & $\surd$  & $\surd$  & $\surd$ & $\surd$ &	$\surd$  & $\surd$  &   --  \\
\object{HD 100453}                 &  I   & $\surd$ &  --      &  --      & $\surd$ & $\surd$ &	$\surd^{ c}$& $\surd$  &   --  \\
\object{HD 100546}                 &  I   & $\surd$ & $\surd^{ a}$&  --      & $\surd$ & $\surd$ &	$\surd$  & $\surd$  &  $E$  \\
\object{HD 135344}     		  &  I   &  --     &  --      &  --      & $\surd^{ b}$ &  --     &	 --      &  --      &   --  \\
\object{HD 139614}                 &  I   &  --     &  --      &  --      & $\surd$ & $\surd$ &	$\surd$  & $\surd$  &   --  \\
\object{HD 142527}                 &  I   & $\surd$ &  --      &  --      &  --     &  --     &	 --      & $\surd$  &  $E$  \\
\object{CD$-$42$^\circ$11721} 		  &  I   & $\surd$ &  --      &  --      & $\surd$ & $\surd$ &	$\surd$  & $\surd$  &   --  \\
\object{HD 169142}                 &  I   & $\surd^{ d}$ &  --      &  --      & $\surd$ & $\surd$ &	$\surd$  & $\surd$  &   --  \\
\object{T CrA}         		  &  I   &  --     &  --      &  --      &  --     &  --     &	 --      &  --      &  $E$  \\
\object{HD 179218}     		  &  I   & $\surd$ &  --      &  --      & $\surd$ & $\surd$ &	$\surd$  & $\surd$  &  $E$  \\
\object{BD+40$^\circ$4124}    	  &  I?  & $\surd^{ d}$ & $\surd$  & $\surd$  & $\surd$ & $\surd$ &	$\surd$  & $\surd$  &   -- \\
\object{HD 200775}     		  &  I   &  --     &  --      &  --      & $\surd$ &  --     &	 --      &  --      &  $E$  \\
\object{MWC 1080}      		  &  I   & $\surd$ &  --      &  --      & $\surd$ & $\surd$ &	$\surd$  & $\surd$  &   --  \\
\hline	            	               	               	  
\object{VX Cas}        		  &  II  & $\surd$ &  --      &  --      & $\surd$ & $\surd$ &	$\surd$  &  --      &  $E$  \\
\object{HD 31648}      		  &  II  &  --     &  --      &  --      & $\surd$ &  --     &	 --      &  --      &  $E$  \\
\object{UX Ori}        		  &  II  &  --     &  --      &  --      &  --     &  --     &	 --      &  --      &  $E$  \\
\object{HD 35187}      		  &  II  &  --     &  --      &  --      &  --     &  --     &	 --      &  --      &   --  \\
\object{BF Ori}        		  &  II  &  --     &  --      &  --      &  --     &  --     &	 --      &  --      &  $E$  \\
\object{RR Tau}        		  &  II  &  --     &  --      &  --      & $\surd$ & $\surd$ &	$\surd$  & $\surd$  &   --  \\
\object{HD 95881}      		  &  II  &  --     &  --      &  --      & $\surd$ & $\surd$ &  $\surd$  & $\surd$  &   --  \\
\object{HD 104237}     		  &  II  &  --     &  --      &  --      &  --     &  --     &	 --      &  --      &  $E$  \\
\object{HD 141569}     		  &  II  &  --     &  --      &  --      & $\surd$ & $\surd$ &	$\surd$  &  --      &   --  \\
\object{HD 142666}     		  &  II  & $\surd$ &  --      &  --      & $\surd$ &  --     &	 --      & $\surd$  &  $E$  \\
\object{HD 144432}                 &  II  &  --     &  --      &  -- &$\surd^{ b}$ &  --     &   --      & $\surd^{ c}$  &  $E$  \\ 
\object{HR 5999}       		  &  II  &  --     &  --      &  --      &  --     &  --     &	 --      & $\surd$  &  $E$  \\
\object{Wra 15-1484}   		  &  II  & $\surd$ &  --      &  --      & $\surd$ & $\surd$ &	$\surd$  & $\surd$  &   --  \\
\object{HD 150193}                 &  II  &  --     &  --      &  --      &  --     &  --     &	 --      & $\surd$  &  $E$  \\
\object{AK Sco}        		  &  II  &  --     &  --      &  --      &  --     &  --     &	 --      &  --      &  $E$  \\
\object{HD 163296}                 &  II  &  --     &  --      &  --      & $\surd$ &  --     &	 --      & $\surd$  &  $E$  \\
\object{VV Ser}        		  &  II  &  --     &  --      &  --      &$\surd^{ b}$ &  --     &	 --      & $\surd$  &  $E$  \\
\object{R CrA}         		  &  II? &  --     &  --      &  --      &  --     &  --     &	 --      &  --      &  $E$  \\
\object{WW Vul}        		  &  II  &  --     &  --      &  --      &  --     &  --     &	 --      &  --      &  $E$  \\
\object{LkH$\alpha$ 224}  	  &  II? &  --     &  --      &  --      &  --     &  --     &	 --      &  --      &   --  \\
\object{BD+46$^\circ$3471}    		  &  II  & $n.s.$  & $n.s.$   & $n.s.$   &  --     &  --     &	 --      &  --      &  $E$  \\
\object{SV Cep}        		  &  II  &  --     &  --      &  --      &  --     &  --     &	 --      &  --      &  $E$  \\
\hline		             	               	               	  
\object{Z CMa}         		  &  III &  --     &  --      &  --      &  --     &  --     &	 --      &  --      &  $A$  \\
\object{IRAS 12496-7650}		  &  III &  --     &  --      &  --      &  --     &  --     &	 --      &  --      &  $A$  \\
\object{MWC 297}       		  &  III & $\surd$ &  --      & $\surd$  & $\surd$ & $\surd$ &	 --      & $\surd$  &  $A$  \\
\object{LkH$\alpha$ 225}  	  &  III &  --     &  --      &  --      &  --     &  --     &	 --      &  --      &  $A$  \\
\object{PV Cep}        		  &  III &  --     &  --      &  --      &  --     &  --     &	 --      &  --      &  $A$  \\
\object{V645 Cyg}      		  &  III &  --     &  --      &  --      &  --     &  --     &	 --      &  --      &  $A$  \\   
 \hline			  
\end{tabular}		  
\end{center}		    
\end{table*}		

\begin{table*}
\caption{ The line fluxes LF of the IR features included in this
analysis. $a(-b)$ represents $a \times 10^{-b}$.
 \label{LF}}	
\scriptsize
\begin{center}
\begin{tabular}{|c|c|r|r|r|r|r|r|r|r|}
 \hline
 \multicolumn{10}{|c|}{\bf LF of the IR features.}\\ \hline \hline
 \multicolumn{1}{|c|}{Object} &
 \multicolumn{1}{|c|}{Group} &
 \multicolumn{1}{|c|}{PAH 3.3} & 
 \multicolumn{1}{|c|}{NAN 3.4} & 
 \multicolumn{1}{|c|}{NAN 3.5} &
 \multicolumn{1}{|c|}{PAH 6.2} &
 \multicolumn{1}{|c|}{PAH 7.7} &
 \multicolumn{1}{|c|}{PAH 8.6} &
 \multicolumn{1}{|c|}{COMP 11} &
 \multicolumn{1}{|c|}{Si 9.7} \\ 
 \multicolumn{1}{|c|}{} &
 \multicolumn{1}{|c|}{} &
 \multicolumn{1}{|c|}{[W/m$^2$]} & 
 \multicolumn{1}{|c|}{[W/m$^2$]} & 
 \multicolumn{1}{|c|}{[W/m$^2$]} &
 \multicolumn{1}{|c|}{[W/m$^2$]} &
 \multicolumn{1}{|c|}{[W/m$^2$]} &
 \multicolumn{1}{|c|}{[W/m$^2$]} &
 \multicolumn{1}{|c|}{[W/m$^2$]} &
 \multicolumn{1}{|c|}{[W/m$^2$]} \\ 
\hline \hline 
\object{V376 Cas}    & I  &  $<$2.38(-15) &  $<$2.82(-15)   &   $<$2.26(-15)   &   $<$1.62(-14) &  $<$2.34(-14) &  $<$9.64(-15)   &  $<$8.88(-15) &     5.39(-13)  \\
\object{Elias 3-1}   & I  &     7.28(-15) &     1.13(-14)   &      1.38(-14)   &      4.49(-14) &     1.23(-13) &     3.71(-14)   &     7.40(-14) &     9.11(-13)  \\
\object{AB Aur}      & I  &  $<$5.33(-15) &  $<$3.26(-15)   &   $<$2.61(-15)   &      3.99(-14) &     5.63(-14) &     3.68(-14)   &     4.61(-14) &     1.47(-12)  \\
\object{HD 34282}    & I  &  $<$2.81(-15) &  $<$3.03(-15)   &   $<$2.43(-15)   &   $<$2.43(-14) &     2.74(-14) &  $<$1.16(-14)   &  $<$1.08(-14) &  $<$1.06(-13)  \\
\object{HD 34700}    & I  &  $<$2.81(-15) &  $<$3.45(-15)   &   $<$2.49(-15)   &      8.75(-15) &     1.46(-14) &     3.28(-15)   &  $<$1.28(-15) &  $<$9.74(-15)  \\
\object{HD 97048}    & I  &     1.85(-14) &     4.71(-14)   &      8.03(-14)   &      8.83(-14) &     2.40(-13) &     4.52(-14)   &     3.97(-14) &  $<$8.95(-15)  \\
\object{HD 100453}   & I  &     8.99(-15) &  $<$3.86(-15)   &   $<$3.10(-15)   &      2.69(-14) &     2.70(-14) &  $<$1.65(-14)   &     7.83(-15) &  $<$3.34(-14)  \\
\object{HD 100546}   & I  &     2.40(-14) &     8.80(-15)   &   $<$1.74(-15)   &      1.21(-13) &     3.22(-13) &     5.61(-14)   &     3.89(-13) &     3.18(-12)  \\
\object{HD 135344}   & I  &  $<$3.06(-15) &  $<$3.03(-15)   &   $<$2.43(-15)   &      1.09(-14) &  $<$9.16(-15) &  $<$3.61(-15)   &  $<$1.56(-15) &  $<$2.41(-14)  \\
\object{HD 139614}   & I  &  $<$2.83(-15) &  $<$3.22(-15)   &   $<$2.59(-15)   &      6.07(-15) &     1.40(-14) &     8.32(-15)   &     1.23(-15) &  $<$2.91(-14)  \\
\object{HD 142527}   & I  &     7.21(-15) &  $<$3.56(-15)   &   $<$2.85(-15)   &   $<$1.69(-14) &  $<$2.41(-14) &  $<$9.90(-15)   &     2.09(-14) &     5.52(-13)  \\
\object{CD$-$42$^\circ$11721} & I  &     6.87(-14) &  $<$4.66(-15)   &   $<$3.73(-15)   &      4.89(-13) &     1.34(-12) &     4.13(-13)   &     3.54(-13) &  $<$9.08(-14)  \\
\object{HD 169142}   & I  &  $<$7.38(-15) &  $<$2.96(-15)   &   $<$2.37(-15)   &      1.93(-14) &     5.68(-14) &     2.55(-14)   &     8.18(-15) &  $<$2.19(-14)  \\
\object{T CrA}       & I  &  $<$1.43(-15) &  $<$1.72(-15)   &   $<$1.38(-15)   &   $<$3.10(-15) &  $<$7.17(-15) &  $<$3.75(-15)   &  $<$3.24(-15) &     1.92(-13)  \\
\object{HD 179218}   & I  &     1.52(-14) &  $<$2.53(-15)   &   $<$2.03(-15)   &      7.07(-14) &     1.42(-13) &     2.66(-14)   &     1.39(-13) &     1.03(-12)  \\
\object{BD+40$^\circ$4124}  & I? &  $<$1.20(-14) &     4.10(-15)   &      4.13(-15)   &      5.38(-14) &     9.84(-14) &     3.32(-14)   &     5.37(-14) &  $<$4.70(-14)  \\
\object{HD 200775}   & I  &  $<$3.08(-15) &  $<$3.25(-15)   &   $<$2.60(-15)   &      3.35(-14) &  $<$2.05(-14) &  $<$9.12(-15)   &  $<$1.19(-14) &     3.36(-13)  \\
\object{MWC 1080}    & I  &     1.14(-14) &  $<$3.84(-15)   &   $<$3.08(-15)   &      3.64(-14) &     9.09(-14) &     2.90(-14)   &     5.51(-14) &  $<$1.16(-13)  \\
\hline               	   		   		      			 		 		 		   		   
\object{VX Cas}      & II &     3.00(-15) &  $<$2.14(-15)   &   $<$1.50(-15)   &      3.17(-15) &     6.25(-15) &     2.71(-15)   &  $<$7.28(-16) &     3.91(-14)  \\
\object{HD 31648}    & II &  $<$2.41(-15) &  $<$3.25(-15)   &   $<$2.61(-15)   &      1.59(-14) &  $<$2.19(-14) &  $<$1.02(-14)   &  $<$9.59(-15) &     5.85(-13)  \\
\object{UX Ori}      & II &  $<$1.41(-15) &  $<$1.76(-15)   &   $<$1.23(-15)   &   $<$6.53(-16) &  $<$2.32(-15) &  $<$1.57(-15)   &  $<$7.20(-16) &     1.32(-13)  \\
\object{HD 35187}    & II &  $<$5.41(-15) &  $<$3.35(-15)   &   $<$2.69(-15)   &   $<$1.41(-14) &  $<$2.36(-14) &  $<$1.06(-14)   &  $<$1.15(-14) &  $<$1.09(-13)  \\
\object{BF Ori}      & II &  $<$1.39(-15) &  $<$1.74(-15)   &   $<$1.21(-15)   &   $<$7.38(-16) &  $<$2.35(-15) &  $<$1.26(-15)   &  $<$6.10(-16) &     5.11(-14)  \\
\object{RR Tau}      & II &  $<$1.62(-15) &  $<$2.01(-15)   &   $<$1.41(-15)   &      1.07(-14) &     2.38(-14) &     1.01(-14)   &     2.53(-15) &  $<$5.43(-15)  \\
\object{HD 95881}    & II &  $<$3.58(-15) &  $<$2.32(-15)   &   $<$1.86(-15)   &      2.15(-14) &     1.96(-14) &     8.12(-15)   &     4.05(-15) &  $<$4.12(-14)  \\
\object{HD 104237}   & II &  $<$2.54(-15) &  $<$2.98(-15)   &   $<$2.39(-15)   &   $<$2.00(-15) &  $<$6.52(-15) &  $<$2.67(-15)   &  $<$1.56(-15) &     7.32(-13)  \\
\object{HD 141569}   & II &  $<$2.42(-15) &  $<$3.01(-15)   &   $<$2.11(-15)   &      6.91(-15) &     1.62(-14) &     2.94(-15)   &  $<$8.02(-16) &  $<$7.81(-15)  \\
\object{HD 142666}   & II &     2.17(-15) &  $<$1.73(-15)   &   $<$1.38(-15)   &      2.37(-14) &  $<$1.30(-14) &  $<$5.37(-15)   &     5.42(-15) &     3.85(-13)  \\
\object{HD 144432}   & II &  $<$1.75(-15) &  $<$1.94(-15)   &   $<$1.55(-15)   &      8.98(-15) &  $<$1.15(-14) &  $<$4.75(-15)   &     3.54(-15) &     3.88(-13)  \\
\object{HR 5999}     & II &  $<$4.65(-15) &  $<$4.55(-15)   &   $<$3.65(-15)   &   $<$7.54(-15) &  $<$1.74(-14) &  $<$7.17(-15)   &     9.95(-15) &     2.78(-13)  \\
\object{Wra 15-1484} & II &     5.57(-15) &  $<$4.60(-15)   &   $<$3.69(-15)   &      4.76(-14) &     1.16(-13) &     2.43(-14)   &     2.36(-14) &  $<$8.36(-14)  \\
\object{HD 150193}   & II &  $<$3.06(-15) &  $<$3.80(-15)   &   $<$2.66(-15)   &   $<$1.81(-15) &  $<$4.96(-15) &  $<$3.64(-15)   &     8.29(-15) &     1.12(-12)  \\
\object{AK Sco}      & II &  $<$2.90(-15) &  $<$3.60(-15)   &   $<$2.52(-15)   &   $<$1.65(-15) &  $<$5.40(-15) &  $<$2.22(-15)   &  $<$1.13(-15) &     7.88(-14)  \\
\object{HD 163296}   & II &  $<$3.04(-15) &  $<$3.15(-15)   &   $<$2.53(-15)   &      2.61(-14) &  $<$1.35(-14) &  $<$5.57(-15)   &     1.49(-14) &     8.41(-13)  \\
\object{VV Ser}      & II &  $<$2.63(-15) &  $<$3.27(-15)   &   $<$2.29(-15)   &      1.06(-14) &  $<$3.30(-15) &  $<$1.67(-15)   &     2.04(-15) &     8.83(-14)  \\
\object{R CrA}       & II?&  $<$8.35(-15) &  $<$9.63(-15)   &   $<$7.72(-15)   &   $<$1.85(-14) &  $<$4.58(-14) &  $<$2.25(-14)   &  $<$1.28(-14) &     1.67(-12)  \\
\object{WW Vul}      & II &  $<$2.00(-15) &  $<$2.49(-15)   &   $<$1.74(-15)   &   $<$1.43(-15) &  $<$3.69(-15) &  $<$1.51(-15)   &  $<$1.09(-15) &     1.27(-13)  \\
\object{LkH$\alpha$ 224}& II?&  $<$3.14(-15) &  $<$3.59(-15)   &   $<$2.88(-15)   &   $<$1.42(-14) &  $<$2.41(-14) &  $<$1.13(-14)   &  $<$1.01(-14) &  $<$9.30(-14)  \\
\object{BD+46$^\circ$3471}  & II &                   &                      &               &   $<$5.90(-15) &  $<$1.42(-14) &  $<$7.28(-15)   &  $<$4.75(-15) &     3.19(-14)  \\
\object{SV Cep}      & II &  $<$1.67(-15) &  $<$2.19(-15)   &   $<$1.76(-15)   &   $<$7.58(-16) &  $<$2.93(-15) &  $<$1.20(-15)   &  $<$9.75(-16) &     2.35(-13)  \\
\hline			   		   		      			 		 		 		   		   
\object{Z CMa}       &III &  $<$6.98(-15) &  $<$8.29(-15)   &   $<$6.65(-15)   &   $<$4.86(-14) &  $<$1.22(-13) &  $<$5.48(-14)   &  $<$4.60(-14) &    -3.50(-12)  \\
\object{IRAS12496-7650}   &III &  $<$1.09(-14) &  $<$1.35(-14)   &   $<$9.50(-15)   &   $<$5.41(-15) &  $<$1.51(-14) &  $<$6.21(-15)   &  $<$2.75(-15) &    -4.98(-13)  \\
\object{MWC 297}     &III &     1.93(-14) &  $<$1.06(-14)   &      2.73(-14)   &      1.54(-13) &     1.89(-13) &  $<$2.98(-14)   &     4.88(-14) &    -9.47(-13)  \\
\object{LkH$\alpha$ 225}&III &  $<$2.25(-15) &  $<$2.64(-15)   &   $<$2.12(-15)   &   $<$1.15(-14) &  $<$1.77(-14) &  $<$7.97(-15)   &  $<$9.30(-15) &    -1.80(-12)  \\
\object{PV Cep}      &III &  $<$2.67(-15) &  $<$3.05(-15)   &   $<$2.45(-15)   &   $<$2.76(-15) &  $<$1.10(-14) &  $<$4.52(-15)   &  $<$2.24(-15) &    -4.99(-13)  \\
\object{V645 Cyg}    &III &  $<$3.65(-15) &  $<$4.54(-15)   &   $<$3.64(-15)   &   $<$2.00(-14) &  $<$3.64(-14) &  $<$1.63(-14)   &  $<$2.04(-14) &    -2.33(-12)  \\
 \hline			  
\end{tabular}		  
\end{center}		    
\normalsize
\end{table*}


\begin{table*}[tbh!]
\caption{ The continuum flux at the peak wavelength of the IR
features included in this analysis. When the amorphous 10 micron
silicate band was present, the PAH features are superimposed on the
band. In this case (flagged by $^\ast$), the silicate feature is
regarded as being part of the underlying 
continuum of the PAH features. Hence the indicated continuum flux
of the PAH band is the flux of the silicate feature at the central
wavelength of that PAH band.
 \label{CF}}	
\begin{center}
\begin{tabular}{|c|c|c|c|c|c|c|c|c|c|}
 \hline
 \multicolumn{10}{|c|}{\bf CF at the peak wavelength of the IR features.}\\ \hline \hline
 \multicolumn{1}{|c|}{Object} &
 \multicolumn{1}{|c|}{Group} &
 \multicolumn{1}{|c|}{PAH 3.3} & 
 \multicolumn{1}{|c|}{NAN 3.4} & 
 \multicolumn{1}{|c|}{NAN 3.5} &
 \multicolumn{1}{|c|}{PAH 6.2} &
 \multicolumn{1}{|c|}{PAH 7.7} &
 \multicolumn{1}{|c|}{PAH 8.6} &
 \multicolumn{1}{|c|}{COMP 11} &
 \multicolumn{1}{|c|}{Si 9.7} \\ 
                 &    &  [Jy]  &  [Jy]  &  [Jy] &  [Jy] &  [Jy]      &  [Jy] &  [Jy] &  [Jy] \\
\hline \hline 
\object{V376 Cas}         & I  &   5.5  &   6.0  &   6.3 &  10.0 &  11.3$^\ast$&  17.1$^\ast$&  23.2$^\ast$&  13.7 \\
\object{Elias 3-1}        & I  &   3.1  &   3.3  &   3.5 &   4.0 &   8.0$^\ast$&  12.4$^\ast$&  32.7$^\ast$&  19.1 \\
\object{AB Aur}           & I  &  13.1  &  13.1  &  13.0 &   9.1 &  10.0$^\ast$&  15.1$^\ast$&  24.8$^\ast$&  12.8 \\
\object{HD 34282}         & I  &   0.7  &   0.8  &   0.7 &   1.0 &   0.9      &   0.8      &   0.8      &   0.7 \\
\object{HD 34700}         & I  &   0.5  &   0.5  &   0.3 &   0.4 &   0.7      &   0.7      &   0.8      &   0.7 \\
\object{HD 97048}         & I  &   2.0  &   2.0  &   2.0 &   1.1 &   3.2      &   4.0      &   6.4      &   4.7 \\
\object{HD 100453}        & I  &   4.8  &   4.9  &   4.9 &   3.7 &   4.4      &   4.2      &   6.3      &   4.7 \\
\object{HD 100546}        & I  &   6.0  &   5.7  &   5.7 &   7.0 &  17.3$^\ast$&  28.8$^\ast$&  62.5$^\ast$&  28.4 \\
\object{HD 135344}        & I  &   3.4  &   3.4  &   3.4 &   2.1 &   1.7      &   1.4      &   1.0      &   1.1 \\
\object{HD 139614}        & I  &   1.4  &   1.3  &   1.3 &   1.0 &   1.4      &   1.5      &   2.6      &   2.4 \\
\object{HD 142527}        & I  &   6.4  &   6.4  &   6.4 &   7.0 &   7.7$^\ast$&  10.7$^\ast$&  11.9$^\ast$&   7.1 \\
\object{CD$-$42$^\circ$11721}& I& 23.6  &  25.0  &  25.7 &  41.4 &  44.3      &  43.7      &  45.9      &  39.5 \\
\object{HD 169142}        & I  &   1.8  &   1.7  &   1.7 &   1.7 &   1.6      &   1.6      &   2.0      &   1.6 \\
\object{T CrA}            & I  &   2.2  &   2.2  &   2.3 &   3.0 &   3.3$^\ast$&   4.4$^\ast$&   8.3$^\ast$&   5.9 \\
\object{HD 179218}        & I  &   3.4  &   3.4  &   3.5 &   5.5 &   9.3$^\ast$&  14.2$^\ast$&  18.3$^\ast$&  10.5 \\
\object{BD+40$^\circ$4124} & I? &  5.5  &   5.6  &   5.7 &   7.5 &   9.0      &   9.6      &   8.9      &   8.1 \\
\object{HD 200775}        & I  &  11.8  &  11.7  &  11.8 &   9.4 &   9.8$^\ast$&  11.4$^\ast$&   9.3$^\ast$&   8.2 \\
\object{MWC 1080}         & I  &  11.2  &  11.6  &  11.7 &  14.5 &  14.9      &  14.6      &  16.2      &  13.6 \\
\hline		 			 	 	 	 	 	 		       	       	
\object{VX Cas}           & II &   0.4  &   0.4  &   0.4 &   0.4 &   0.5$^\ast$&   0.8$^\ast$&   1.2$^\ast$&   0.7 \\
\object{HD 31648}         & II &   4.0  &   4.0  &   4.0 &   4.6 &   6.1$^\ast$&   8.7$^\ast$&  11.3$^\ast$&   6.1 \\
\object{UX Ori}           & II &   0.9  &   0.9  &   0.9 &   0.8 &   0.9$^\ast$&   2.0$^\ast$&   2.7$^\ast$&   1.5 \\
\object{HD 35187}         & II &   2.4  &   2.4  &   2.4 &   1.9 &   2.6      &   3.8      &   5.4      &   4.9 \\
\object{BF Ori}           & II &   0.4  &   0.4  &   0.4 &   0.5 &   0.6$^\ast$&   1.1$^\ast$&   1.4$^\ast$&   0.8 \\
\object{RR Tau}           & II &   0.9  &   0.9  &   0.9 &   0.9 &   1.1      &   1.2      &   1.6      &   1.3 \\
\object{HD 95881}         & II &   4.0  &   3.9  &   3.9 &   4.5 &   5.5      &   6.9      &   7.8      &   8.0 \\
\object{HD 104237}        & II &  10.8  &  10.8  &  10.8 &  10.8 &  10.6$^\ast$&  15.9$^\ast$&  22.4$^\ast$&   9.6 \\
\object{HD 141569}        & II &   0.7  &   0.7  &   0.6 &   0.4 &   0.5      &   0.5      &   0.5      &   0.5 \\
\object{HD 142666}        & II &   2.7  &   2.7  &   2.7 &   0.6 &   3.7$^\ast$&   5.5$^\ast$&   6.3$^\ast$&   2.6 \\
\object{HD 144432}        & II &   2.8  &   2.8  &   2.7 &   2.1 &   2.3$^\ast$&   4.8$^\ast$&   7.4$^\ast$&   4.1 \\
\object{HR 5999}          & II &  17.9  &  18.0  &  18.1 &  13.6 &  13.9$^\ast$&  15.0$^\ast$&  16.9$^\ast$&  14.7 \\
\object{Wra 15-1484}      & II &   6.6  &   7.1  &   7.4 &  14.4 &  17.3      &  18.4      &  21.0      &  19.8 \\
\object{HD 150193}        & II &   4.8  &   4.7  &   4.7 &   4.8 &   5.5$^\ast$&  13.3$^\ast$&  18.7$^\ast$&   8.8 \\
\object{AK Sco}           & II &   1.5  &   1.5  &   1.5 &   1.2 &   1.2$^\ast$&   1.9$^\ast$&   2.8$^\ast$&   2.0 \\
\object{HD 163296}        & II &   9.8  &   9.7  &   9.6 &   7.2 &   8.8$^\ast$&  14.0$^\ast$&  20.3$^\ast$&  12.2 \\
\object{VV Ser}           & II &   2.5  &   2.5  &   2.5 &   2.6 &   3.1$^\ast$&   3.8$^\ast$&   4.4$^\ast$&   3.4 \\
\object{R CrA}            & II?&  54.2  &  57.0  &  59.2 &  64.6 &  58.3$^\ast$&  79.8$^\ast$& 126.3$^\ast$&  86.1 \\
\object{WW Vul}           & II &   0.9  &   0.9  &   0.8 &   0.7 &   0.8$^\ast$&   1.5$^\ast$&   2.1$^\ast$&   1.3 \\
\object{LkH$\alpha$ 224}  & II?&   3.9  &   4.1  &   4.2 &   7.3 &   9.0      &  10.3      &  12.4      &  11.4 \\
\object{BD+46$^\circ$3471} & II &       &        &       &   0.6 &   0.9$^\ast$&   1.0$^\ast$&   1.6$^\ast$&   1.0 \\	       
\object{SV Cep}           & II &   0.6  &   0.7	&   0.7	&   0.8 &   1.1$^\ast$&   2.9$^\ast$&   4.6$^\ast$&   2.7 \\       
\hline		 		      	 	 	 	 	 	
\object{Z CMa}            & III&  40.1  &  43.3  &  45.2 &  86.9 & 101.7      &  84.3      & 101.9      & 114.7 \\
\object{IRAS 12496-7650}	 & III&   7.9  &   9.1  &   9.7 &  17.7 &  21.8      &  21.8      &  33.3      &  30.2 \\
\object{MWC 297}          & III&  65.8  &  68.3  &  71.5 & 104.6 & 118.9      & 123.7      & 119.7      & 123.3 \\
\object{LkH$\alpha$ 225}  & III&   1.3  &   1.5  &   1.7 &  12.6 &  24.5      &  23.8      &  26.2      &  36.8 \\
\object{PV Cep}           & III&   1.9  &   2.2  &   2.4 &   7.4 &   9.6      &   8.5      &  10.6      &  12.8 \\
\object{V645 Cyg}         & III&   8.3  &   9.6  &  10.7 &  45.2 &  65.7      &  65.4      &  87.5      &  91.0 \\
\hline
\end{tabular}		  
\end{center}		    
\end{table*}

\subsection{The SED \label{thesed}}

To characterize the spectral energy distribution (SED) of the sample
sources, several 
quantities were determined, based on UV to millimetre (mm) photometry from the
literature. The photometry consists of ANS and IUE ultraviolet data, 
ground-based photometry in the Walraven, Johnson/Cousins, and near-IR 
$JHKLM$ photometric systems, IRAS and MSX data, and single-dish (sub)-mm 
photometry collected from the literature. The effective temperature 
$T_{\eff}$ of each source was determined from its spectral type listed in 
Table~\ref{parameters}, using the calibrations from \citet{schmidtkaler}.
The visual extinction $A_V$ and absolute luminosity $L$ of the central 
star were computed by fitting and integrating a \citet{kurucz} model for 
the stellar photosphere to the de-reddened photometry, adopting the 
distance estimates listed in Table~\ref{parameters}. 
The ionising ($>$13.6 eV) luminosity $L_{\ion}$ and the 
UV (2-13.6 eV) luminosity $L_{\uv}$
of the source were computed using this Kurucz model fit. We also computed 
the IR-excess 
luminosity $L_{\exc}$, using a spline fit to the infrared data,
and the absorbed luminosity $L_{\abs}$ of the source, which is the 
difference between the theoretical Kurucz model and the reddened model. The
IR-excess flux in the K band (2.2 $\mu$m), at 60, 850, and 1300 micron
was deduced by subtracting the
Kurucz model at these wavelengths from the observed fluxes. When no
1300 micron photometry was available, we included 1100 micron
measurements, which we then converted into 1300 micron
flux values by multiplying with the average 1300/1100 ratio deduced
from stars for which we had both measurements.
The last parameter used to characterize the SED is the observed bolometric
luminosity $L_{\bol}$, which is the total integrated luminosity of the
SED. The latter quantity is \textit{not} corrected for extinction, as
opposed to the stellar luminosity $L$. The stellar parameters used in
this study are listed in Table~\ref{parameters}.

\begin{figure}
\rotatebox{0}{\resizebox{3.5in}{!}{\includegraphics{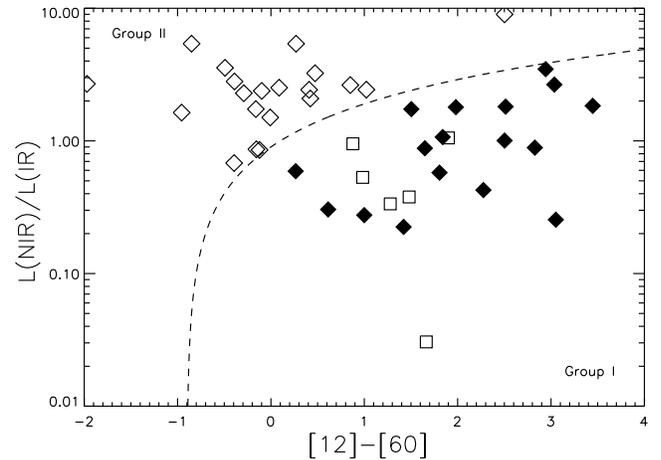}}}
\caption{ Diagram based on \citet{vanboekel}. The ratio
$L_{\mathrm{NIR}}/L_{\mathrm{IR}}$ is plotted versus the non-color-corrected IRAS
[12]-[60] color. The dashed line represents $L_{\mathrm{NIR}}/L_{\mathrm{IR}} =
([12]-[60]) + 0.9$. Objects on the left-hand side of this empirical
line are defined to be group II members (open diamonds), sources on
the right side are either group I (filled diamonds) or group III
(squares) members.}
\label{plotroy.ps}
\end{figure}

The sample sources were classified into different groups, based on the shape
of their SED. The quantities used to
characterize the infrared spectral energy distribution of HAEBEs are
the ratio of $L_{\mathrm{NIR}}$ (the integrated luminosity derived from the
broad-band $J$, $H$, $K$, $L$ and $M$ photometry) and
$L_{\mathrm{IR}}$ (the corresponding 
quantity derived from IRAS 12, 25 and 60 micron points), and the non-color-corrected 
IRAS [12]-[60] color
\citep{vanboekel,dullemond03,ackesubmm}. These parameters compare the
near-IR SED, which is expected to be similar in HAEBEs
\citep{hillenbrand, dewinter95, natta01}, to the mid-IR SED, where 
the major differences in SED shape occur. 
M01 Group I sources are stronger mid-IR emitters than group II sources.
The luminosity ratio
$L_{\mathrm{NIR}}/L_{\mathrm{IR}}$ represents the {\it strength} of the
near-IR compared to the mid-IR excess, which is lower for group I than
group II sources. The shape of the mid-IR SED of a group I source is
``double-peaked'' compared to the SED of a group II member. The $IRAS$
[12]-[60] color index provides a quantitative measure for this
difference in SED {\it shape}. Group I sources are redder than their group
II counterparts. 
We use the diagram given in
Fig.~\ref{plotroy.ps} to distinguish between group I and group II in
the classification of M01. The dashed line represents
$L_{\mathrm{NIR}}/L_{\mathrm{IR}} = ([12]-[60]) + 0.9$, which
empirically provides the best separation between the two groups. 
 
Six of our sample stars display the amorphous 10 micron feature in
absorption. These objects are believed to possess disks whose
luminosity is dominated by viscous dissipation of energy due to
accretion, are deeply embedded systems and hence are fundamentally
different from the other sample stars. We therefore classify them in a
different group: group III.

\object{BD+40$^\circ$4124}, \object{R CrA} and \object{LkH$\alpha$ 224} have
not been classified 
based on their appearance in Fig.~\ref{plotroy.ps}. Confusion with
background sources in the IRAS photometry prohibited us from deriving the
quantities needed to plot these objects in the
diagram. \object{BD+40$^\circ$4124} has been classified as a group I source,
because its SED resembles the SED of \object{HD 200775}. \object{R CrA}
and \object{LkH$\alpha$ 224} are both \hbox{UX Orionis} stars
according to the definition of \citet{dullemond03}: both sources have
a spectral type later than B9 and 
optical variations larger than 1 mag on timescales of days to weeks. 
Therefore we classified them as group II members.

\begin{figure*}
\rotatebox{0}{\resizebox{17cm}{!}{\includegraphics{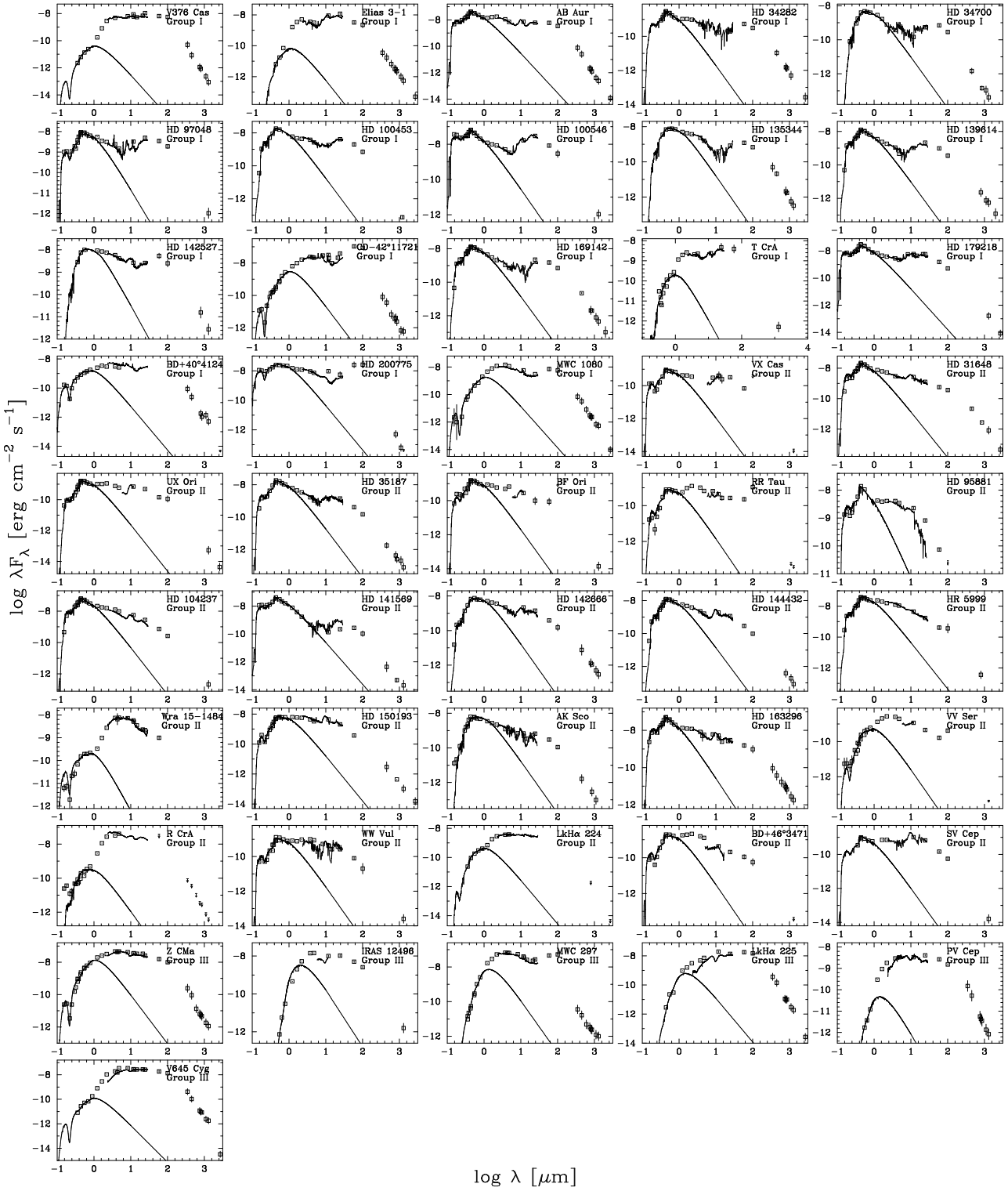}}}
\caption{ The spectral energy distributions of the sample stars. The
  squares are the observed photometric data from the literature with
  error bars. The solid line is the reddened Kurucz model for the
  stellar photosphere, fitted to the 
  measured UV--optical photometry. The ISO spectra are plotted in the
  SEDs as well (noisy solid line).}
\label{seds.ps}
\end{figure*}

The SEDs of all sample sources are displayed in
Fig.~\ref{seds.ps}. The group to which the objects belong is indicated
as well. The SEDs are overplotted with the ISO spectra. Squares with
error bars indicate photometric measurements from the literature,
arrows upper limits.

\begin{sidewaystable*}
\caption{ The stellar parameters for the sample sources. The distance
$d$ which we used in the analysis and spectral type (Sp.T.) are
given. The visual extinction $A_V$, the logarithm of the effective
temperature $\log T_{\eff}$, of the stellar luminosity $\log L$, 
of the stellar UV luminosity $\log L_{\uv}$ and of
the observed bolometric luminosity $\log L_{\bol}$ are presented as
well. The latter quantities are estimated based on the photometric
measurements. Furthermore, the ratio of the IR excess luminosity over
the stellar luminosity is computed. Ten of the sample sources have a
ratio larger than unity, hence are not consistent with a passive
reprocessing disk. The last 4 columns contain the IR excess fluxes in
K (2.2 $\mu$m), at 60, 850 and 1300 micron. $^\bigstar$ These
stars are UX Orionis objects according to the definition of
\citet{dullemond03}. $^{\spadesuit}$ Probable confusion with
the nearby source \object{CrA IRS7}. }
 \label{parameters}
\begin{center}
\begin{tabular}{|c|c|c|c|c|c|c|c|c|c|c|c|c|c|c|c|}
 \hline
 \multicolumn{16}{|c|}{\bf The stellar parameters of the sample stars.}\\ \hline \hline
 \multicolumn{1}{|c|}{Object} &
 \multicolumn{1}{|c|}{Region} & 
 \multicolumn{1}{|c|}{$d$} & 
 \multicolumn{1}{|c|}{Ref.} &
 \multicolumn{1}{|c|}{Sp.T.} &
 \multicolumn{1}{|c|}{Ref.} &
 \multicolumn{1}{|c|}{A$_V$} &
 \multicolumn{1}{|c|}{$\log T_{\eff}$} &
 \multicolumn{1}{|c|}{$\log L$} & 
 \multicolumn{1}{|c|}{$\log L_{\uv}$} & 
 \multicolumn{1}{|c|}{$\log L_{\bol}$} &  
 \multicolumn{1}{|c|}{$L_{\exc}/L$} &
 \multicolumn{1}{|c|}{$\Delta$ K} &
 \multicolumn{1}{|c|}{$\Delta$ 60} & 
 \multicolumn{1}{|c|}{$\Delta$ 850} & 
 \multicolumn{1}{|c|}{$\Delta$ 1300} \\ 
               &            & [pc]  &       &      &      &[mag] &$\log$[K] &$\log [L_{\odot}]$& $\log [L_{\odot}]$& $\log [L_{\odot}]$&   & [mag] & [mag] & [mag] & [mag]   \\
\hline \hline	 
group I &  &  &  &  &  &  &  &  &  &  &  &  &  &   & \\
\hline
\object{V376 Cas}       & Cas R1     &  630  &  S89  &  B5e        &  C79 & 4.40   &  4.188 &   1.63  &   1.60   &   2.57  &    9.35 &  5.65 &     16.45 &     15.48   &    14.61    \\
\object{Elias 3-1}      & L1495      &  160  &  K94  &  A6:e       &  Z94 & 4.05   &  3.904 &  -0.14  &$-$0.33   &   1.35  &   35.69 &  5.56 &     14.89 &     15.92   &    15.78    \\
\object{AB Aur}         & L1519      &  140  &  V98  &  A0Ve+sh    &  B93 & 0.50   &  3.979 &   1.68  &   1.56   &   1.70  &    0.51 &  2.28 &     11.33 &     11.08   &    10.67    \\
\object{HD 34282}$^\bigstar$       & &  400  &  P03  &  A3Vne      &  M00 & 0.28   &  3.941 &   1.27  &   1.11   &   1.37  &    0.53 &  2.05 &     11.68 &     14.04   &     --      \\
\object{HD 34700}       & Orion OB1a &  340  &  D99  &  G0IV+G0IVe &  A03 & 0.00   &  3.774 &   1.31  &   0.91   &   1.42  &    0.27 &  0.08 &     10.21 &     --      &   9.96      \\
\object{HD 97048}       & Ced 111    &  180  &  V98  &  B9.5Ve+sh  &  W87 & 1.26   &  4.000 &   1.64  &   1.53   &   1.41  &    0.35 &  1.23 &     11.52 &     --      &   13.04     \\
\object{HD 100453}      &            &  112  &  HIP  &  A9Ve       &  H75 & 0.02   &  3.869 &   0.90  &   0.66   &   1.09  &    0.57 &  1.27 &     10.71 &     --      &    --       \\
\object{HD 100546}      & Sco OB2-4? &  103  &  V98  &  B9Vne      &  H75 & 0.26   &  4.021 &   1.51  &   1.42   &   1.62  &    0.58 &  1.49 &     11.77 &     --      &    --       \\
\object{HD 135344}      & Sco OB2-3  &  140  &  D99  &  F4Ve       &  D97 & 0.31   &  3.819 &   0.91  &   0.59   &   1.00  &    0.44 &  1.23 &     10.23 &     12.04   &    11.64    \\
\object{HD 139614}      & Sco OB2-3  &  140  &  D99  &  A7Ve       &  D97 & 0.09   &  3.895 &   0.91  &   0.70   &   1.03  &    0.40 &  0.98 &     10.57 &     12.72   &    12.80    \\  
\object{HD 142527}      & Sco OB2-2  &  145  &  D99  &  F7IIIe     &  V98 & 0.64   &  3.796 &   1.18  &   0.82   &   1.34  &    0.82 &  1.28 &     11.08 &     13.47   &    13.18    \\
\object{CD$-$42$^\circ$11721}&       &  400  &  D90  &  B0IVep     &  S90 & 5.08   &  4.470 &   3.94  &   3.92   &   3.02  &    0.13 &  2.83 &     14.87 &     12.19   &    12.07    \\     
\object{HD 169142}      & Sco OB2-1  &  145  &  D99  &  A5Ve       &  D97 & 0.43   &  3.914 &   1.16  &   0.98   &   1.12  &    0.26 &  0.76 &     10.66 &     12.38   &    12.18    \\
\object{T CrA}          & NGC 6729   &  130  &  M81  &  F0e        &  F84 & 2.45   &  3.857 &  -0.18  &$-$0.44   &   0.88  &   12.21 &  3.86 &     14.38 &  $<$13.29   &    14.99    \\
\object{HD 179218}      & L693       &  240  &  V98  &  B9e        &  S66 & 0.54   &  4.021 &   2.00  &   1.90   &   1.88  &    0.28 &  1.27 &     10.53 &      --     &    10.93    \\
\object{BD+40$^\circ$4124}& Cyg R1   &  980  &  S91  &  B2Ve       &  H95 & 3.01   &  4.342 &   3.77  &   3.76   &   2.49  &    0.05 &  2.28 &      --   &     12.72   &    13.34    \\
\object{HD 200775}      & Cep R2     &  440  &  W81  &  B2.5IVe    &  R95 & 1.90   &  4.291 &   3.82  &   3.80   &   2.84  &    0.07 &  1.44 &     12.49 &     9.48    &    8.17     \\
\object{MWC 1080}       & L1238      & 2200  &  L88  &  B0e        &  C79 & 5.27   &  4.477 &   5.26  &   5.24   &   3.75  &    0.03 &  2.98 &     12.41 &     12.60   &    12.44    \\
\hline		
group II &  &  &  &  &  &  &  &  &  & &  &  &  &   & \\
\hline
\object{VX Cas}$^\bigstar$& Cas R1   &  630  &  S89  &  A1Ve+sh   &  G98  & 0.81   &  3.965 &   1.50  &   1.37   &   1.44  &    0.47 &  1.61 &     10.04 &      --     & $<$11.00    \\
\object{HD 31648}       &  	     &  130  &  V98  &  A3Ve      &  J91  & 0.25   &  3.941 &   1.14  &   0.98   &   1.22  &    0.42 &  1.79 &     9.63  &     12.76   &    12.87    \\
\object{UX Ori}$^\bigstar$ &Orion OB1a& 340  & D99   & A4IVe      &  M00  & 1.91   &  3.925 &   1.68  &   1.51   &   1.25  &    0.16 &  0.91 &     8.73  &     --      &   10.49     \\
\object{HD 35187}       &   L1559    & 150   & D98   & A2Ve+A7Ve  &  D98  & 0.71   &  3.953 &   1.44  &   1.29   &   1.23  &    0.15 &  0.99 &     8.91  &     10.20   &    9.99     \\
\object{BF Ori}$^\bigstar$ &   Ori OB1c& 510 & D99   & A2IVev     &  M00  & 0.87   &  3.954 &   1.53  &   1.39   &   1.38  &    0.30 &  1.56 &     9.68  &     --      &   10.51     \\
\object{RR Tau}$^\bigstar$ &   (L1553) & 160 & K94   & A3--5e     &  F84  & 1.04   &  3.927 &   0.31  &   0.14   &   0.42  &    0.96 &  2.73 &     11.06 &      --     & $<$12.15    \\
\object{HD 95881}       &   Sco OB2-4? & 118 & D99   & A2III/IVe  &  H75  & 0.25   &  3.954 &   0.84  &   0.69   &   0.98  &    0.62 &  2.15 &     8.06  &     --      &    --       \\    
\hline
\end{tabular}		  
\end{center}		    
\end{sidewaystable*}		

\begin{sidewaystable*}
\begin{center}
\begin{tabular}{|c|c|c|c|c|c|c|c|c|c|c|c|c|c|c|c|}
 \hline
 \multicolumn{16}{|c|}{\bf The stellar parameters of the sample stars (continued).}\\ \hline \hline
 \multicolumn{1}{|c|}{Object} &
 \multicolumn{1}{|c|}{Region} & 
 \multicolumn{1}{|c|}{$d$} & 
 \multicolumn{1}{|c|}{Ref.} &
 \multicolumn{1}{|c|}{Sp.T.} &
 \multicolumn{1}{|c|}{Ref.} &
 \multicolumn{1}{|c|}{A$_V$} &
 \multicolumn{1}{|c|}{$\log T_{\eff}$} &
 \multicolumn{1}{|c|}{$\log L$} & 
 \multicolumn{1}{|c|}{$\log L_{\uv}$} & 
 \multicolumn{1}{|c|}{$\log L_{\bol}$} &  
 \multicolumn{1}{|c|}{$L_{\exc}/L$} &
 \multicolumn{1}{|c|}{$\Delta$ K} &
 \multicolumn{1}{|c|}{60} & 
 \multicolumn{1}{|c|}{850} & 
 \multicolumn{1}{|c|}{1300} \\ 
               &               & [pc]  &       &      &      &      &$\log$[K] &$\log [L_{\odot}]$&$\log [L_{\odot}]$& $\log [L_{\odot}]$&   & [mag] & [mag] & [mag] & [mag]   \\
\hline \hline	  					                                                                                 
\object{HD 104237}      &   Cha III     & 116   & V98   & A4IVe+sh   &  V98  & 0.29   &  3.925 &   1.54  &   1.36   &   1.53  &    0.23 &  1.18 &     8.54  &     --      &   10.02     \\
\object{HD 141569}      &   (L169)      &  99   & V98   & A0Ve       &  D97  & 0.37   &  3.979 &   1.28  &   1.16   &   1.09  &    0.01 & -0.07 &     8.18  &     11.16   &    8.21     \\
\object{HD 142666}$^\bigstar$&   Sco OB2-2 & 145& D99   & A8Ve       &  D97  & 0.93   &  3.880 &   1.13  &   0.91   &   1.03  &    0.33 &  1.21 &     8.96  &     11.54   &    11.51    \\
\object{HD 144432}      &   Sco OB2-2   & 145   & D99   & A9IVev     &  M00  & 0.17   &  3.866 &   1.01  &   0.76   &   1.13  &    0.45 &  1.40 &     8.89  &     10.41   &    10.37    \\
\object{HR 5999}$^\bigstar$&   Lupus 3  & 210   & V98 & A5--7III/IVe+sh& T89 & 0.49   &  3.899 &   1.94  &   1.74   &   1.97  &    0.46 &  1.63 &     7.98  &     9.08    &  $<$9.19    \\
\object{Wra 15-1484}    &               & 750   & L89   & B0:[e]      & D98b & 3.07   &  4.477 &   2.99  &   2.97   &   2.45  &    0.31 &  4.45 &     13.60 &      --     &     --      \\
\object{HD 150193}      &    Sco OB2-2  & 150   & V98   & A1--3Ve     & G98  & 1.49   &  3.953 &   1.38  &   1.23   &   1.19  &    0.45 &  1.82 &     8.98  &     10.57   &    10.43    \\
\object{AK Sco}$^\bigstar$ &   	        & 150   & V98   & F5+F5IVe    & A89  & 0.62   &  3.809 &   0.95  &   0.62   &   0.88  &    0.21 &  0.64 &     8.69  &     10.07   &     --      \\
\object{HD 163296}      &   	        & 122   & V98   & A3Ve        & G98  & 0.09   &  3.941 &   1.38  &   1.22   &   1.51  &    0.46 &  1.92 &     9.98  &     13.00   &    12.97    \\
\object{VV Ser}$^\bigstar$&    Serpens  & 330   & D91   & A2IIIe      &  V   & 2.67   &  3.954 &   1.27  &   1.13   &   1.36  &    1.37 &  3.13 &     10.08 &      --     & $<$11.39    \\
\object{R CrA}$^\bigstar$&    NGC 6729  & 130   & M81   & A1--F7ev    & V98  & 1.33   &  3.857 &  -0.19  &$-$0.45   &   3.45  & 4675.32$^{\spadesuit}$ &  6.56 &  $<$17.10 &  $<$15.68   & $<$14.89    \\
\object{WW Vul}$^\bigstar$ &    Vul R1  & 440   & V81   & A4IV/Ve+sh  & G98  & 1.18   &  3.979 &   1.44  &   1.31   &   1.29  &    0.50 &  2.19 &     9.71  &      --     &    11.28    \\
\object{LkH$\alpha$ 224}$^\bigstar$&Cyg R1& 980 & S91   & A7e         & V99  & 2.98   &  3.895 &   2.06  &   1.85   &   2.31  &    1.97 &  3.15 &      --   &  $<$14.06   &     --      \\
\object{BD+46$^\circ$3471}&  IC 5146    &1200   & H02   & A0.5IIIe    & G98  & 1.27   &  3.991 &   2.60  &   2.49   &   2.43  &    0.50 &  2.47 &     9.44  &      --     & $<$11.17    \\
\object{SV Cep}$^\bigstar$  &    Cep R2 & 440   & W81   & A0--2Ve     & G98  & 0.68   &  3.965 &   1.15  &   1.01   &   1.39  &    1.29 &  2.49 &     10.95 &      --     &    11.41    \\
\hline		  						  	                                                                                 
group III &  &  &  &  &  &  & &  &  &  &  &  &  &   & \\
\hline
\object{Z CMa}          &    CMa R1    &  1050  &  S99   & Be        &  V98  & 4.46   &  4.477 &   5.15  &   5.12   &   3.71  &    0.05 &  2.57 &     11.88 &     11.98   &    11.85    \\
\object{IRAS12496-7650} &    Cha II    &   180  &  W97   & Ae        &  H91  &14.00   &  3.927 &   2.52  &   2.35   &   1.60  &    0.14 &  0.51 &      9.16 &      --     &    10.71    \\
\object{MWC 297}        &    (L515)    &   250  &  D97c  & B1.5Ve    &  D97c & 7.73   &  4.375 &   4.01  &   4.00   &   2.69  &    0.07 &  2.61 &     12.11 &     10.31   &    10.76    \\
\object{LkH$\alpha$ 225}&    Cyg R1    &   980  &  S91   & A--Fe     &  H95  & 7.32   &  3.857 &   2.62  &   2.36   &   3.25  &    4.64 &  2.05 &     13.41 &     14.19   &    13.79    \\
\object{PV Cep}         &    Cep R2    &   440  &  W81   & A5:e      &  C81  & 7.01   &  3.914 &   0.94  &   0.75   &   1.91  &   11.29 &  4.33 &     14.24 &     15.98   &    15.81    \\
\object{V645 Cyg}       &    Cyg OB7   &  3500  &  G86   & O7e       &  V02  & 4.15   &  4.580 &   4.54  &   4.41   &   4.60  &    1.25 &  6.23 &     16.99 &     17.53   &    17.26    \\
 \hline
 \end{tabular}		  
 \begin{tabular}{|c|c|c|c|c|c|}
  \hline
 \multicolumn{6}{|c|}{References} \\ \hline
 \multicolumn{1}{|c|}{Ref.} &
 \multicolumn{1}{|c|}{Article} &
 \multicolumn{1}{|c|}{Ref.} &
 \multicolumn{1}{|c|}{Article} &
 \multicolumn{1}{|c|}{Ref.} &
 \multicolumn{1}{|c|}{Article} \\
   \hline \hline
A89   & \citet{andersen}                &G98   & \citet{gray}                    &  S89   & \citet{shevchenko89a,shevchenko89b} \\     
A03   & \citet{arellanoferro}           &G99   & \citet{gurtler}                 &  S90   & \citet{shore} \\              
B93   & \citet{bohm93}     	        &H02   & \citet{herbig02}                &  S91   & \citet{shevchenko91} \\         
C79   & \citet{cohen79}     		&H75   & \citet{houk}                    &  S99   & \citet{shevchenko99} \\         
C81   & \citet{cohen81} 		&H91   & \citet{hughes}                  &  T89   & \citet{tjinadjie89} \\        
D90   & \citet{dewinter}        	&H95   & \citet{hillenbrand95}           &  V02   & \citet{valtts} \\        
D91   & \citet{delara}   	        &J91   & \citet{jaschek}                 &  V81   & \citet{voshchinnikov} \\             
D97   & \citet{dunkin97a} 		&K94   & \citet{kenyon}                  &  V98   & \citet{vandenancker98} \\     
D97c  & \citet{drew}    		&L88   & \citet{levreault88}               &  V99   & \citet{vandenancker99} \\     
D98   & \citet{dunkin98}        	&L89   & \citet{lebertre}                &   V    & van den Ancker (unpublished) \\    
D98b  & \citet{dewinter98}      	&M81   & \citet{marraco}                 &  W81   & \citet{whitcomb} \\           
D99   & \citet{dezeeuw}         	&M00   & \citet{merin}                   &  W87   & \citet{whittet87} \\            
F84   & \citet{finkenzeller84}   	&P03   & \citet{pietu}                   &  W97   & \citet{whittet97} \\            
F85   & \citet{finkenzeller85} 		&R95   & \citet{rogers}                  &  Z94   & \citet{zinnecker} \\    
G86   & \citet{goodrich} 		&S66   & \citet{slettebak}               &  HIP   & Hipparcos  \\
  \hline 
\end{tabular}		  
\end{center}		    
\end{sidewaystable*}		


\section{Analysis of the ISO spectra}

\subsection{The PAH features \label{thePAHfeatures}}

\begin{figure}
\rotatebox{0}{\resizebox{3.5in}{!}{\includegraphics{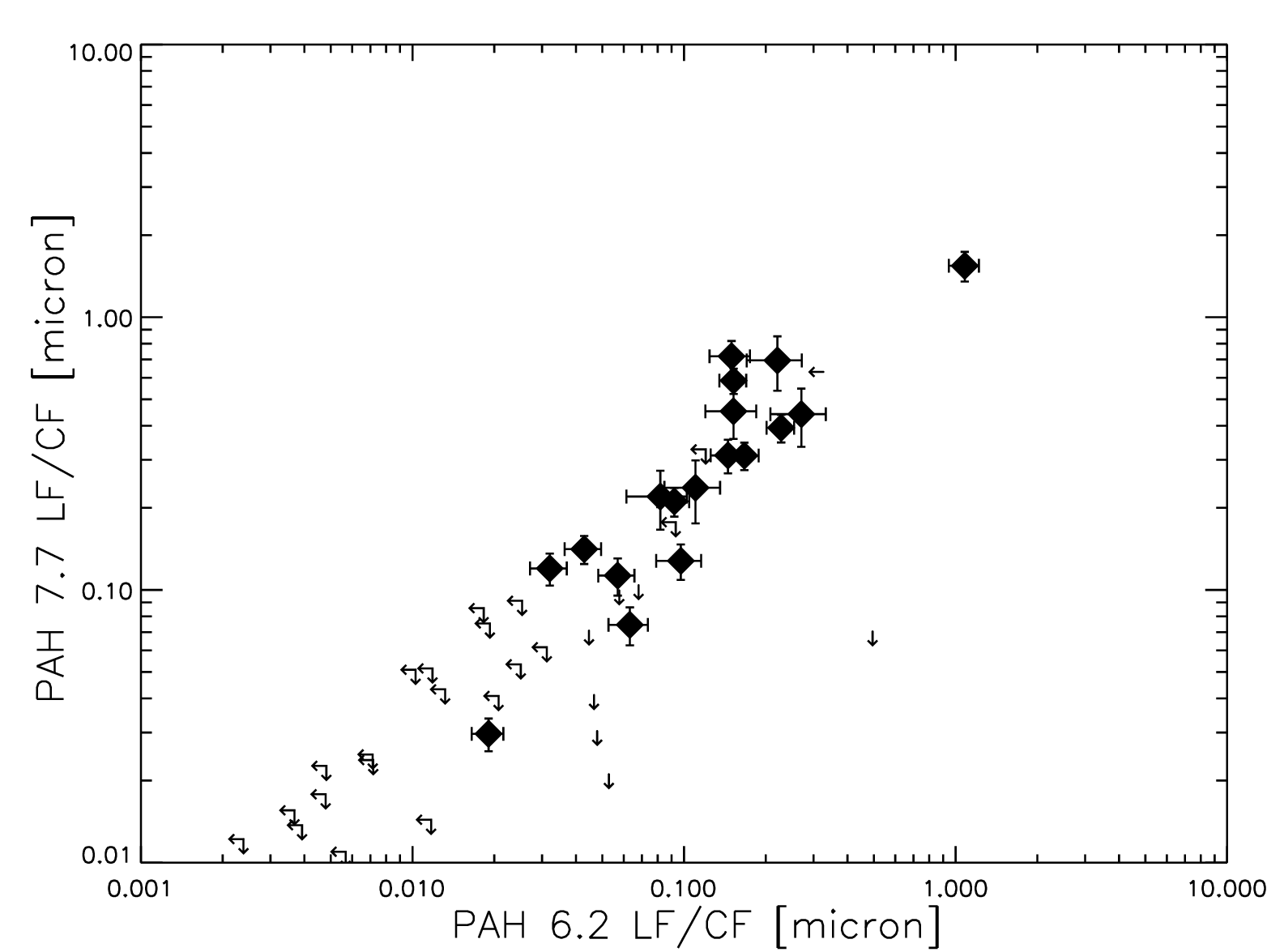}}}
\caption{ The $LF/CF$ ratio of PAH 7.7 micron versus PAH 6.2
micron. The filled diamonds with error bars indicate detected features,
arrows represent upper limits.}
\label{LFoverCF6277.ps}
\end{figure}

\begin{figure}
\rotatebox{0}{\resizebox{3.5in}{!}{\includegraphics{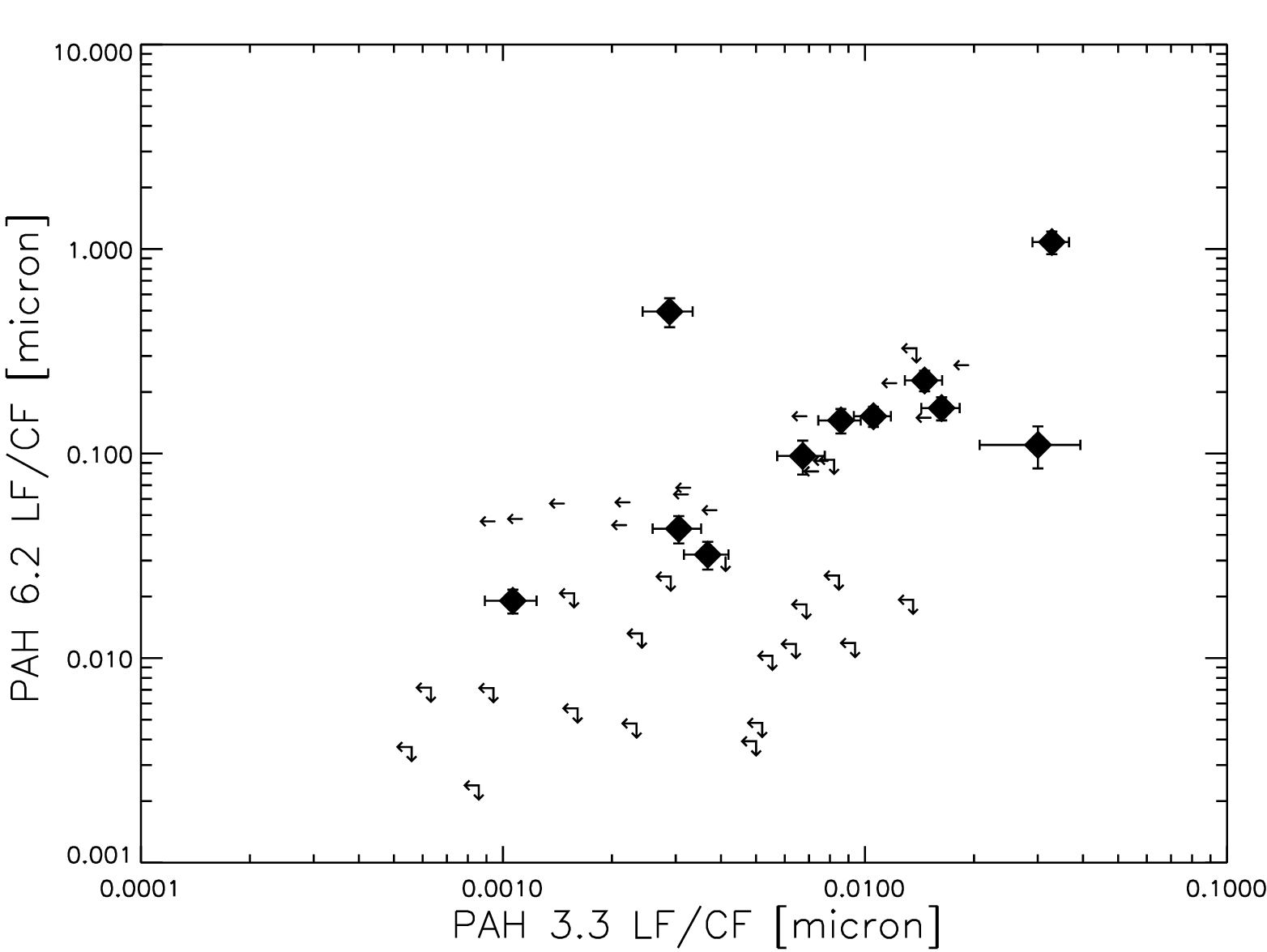}}}
\caption{ Similar plot as Fig.~\ref{LFoverCF6277.ps}. The $LF/CF$
ratio of PAH 6.2 micron versus PAH 3.3 micron.}
\label{LFoverCF3362.ps}
\end{figure}

First we investigated the general correlations of the different PAH
features in our spectrum. We compared the line fluxes (LF) of the
different features, which we normalised by the continuum flux at the
peak wavelength (CF), to correct for the
distance-to-the-star dependence of the line flux. 
Fig.~\ref{LFoverCF6277.ps} shows the ratio $LF/CF$ for the PAH 7.7
micron feature versus PAH 6.2 micron; Fig.~\ref{LFoverCF3362.ps} is a
similar plot for the ratio $LF/CF$ of PAH 6.2 micron versus PAH 3.3
micron. The correlation coefficient $\eta$ in the first plot is 0.90,
while the data in the second plot are less correlated ($\eta =
0.58$). The other PAH features display comparable behaviour. It is
striking though that PAH features that are linked to the CC bonds
(6.2, 7.7 micron) correlate better with each other, as well as the
features linked to the CH bonds (3.3, 7.7, 8.6 micron). 
However, as a first order approximation we can state that the $LF/CF$
ratios correlate fairly well over a range of a few orders of
magnitude.

\begin{figure}
\rotatebox{0}{\resizebox{3.5in}{!}{\includegraphics{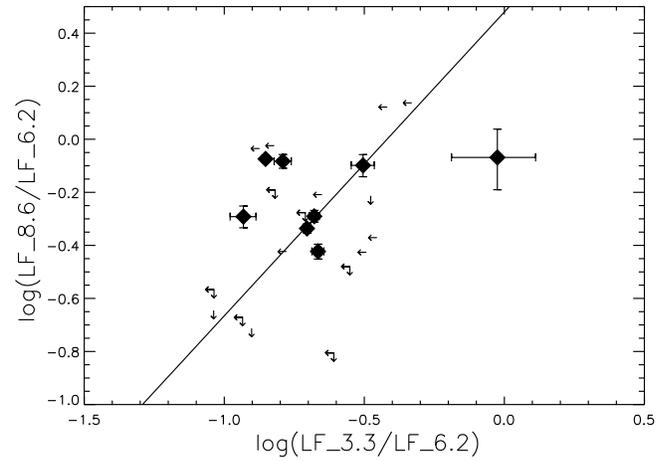}}}
\caption{ The logarithm of the ratio of the line fluxes of the PAH
features at 8.6 and 6.2 micron
$\log(LF_{8.6}/LF_{6.2})$ versus the logarithm of the ratio of the
line fluxes of the PAH features at 3.3 and 6.2 micron
$\log(LF_{3.3}/LF_{6.2})$ for all program stars. The filled diamonds
with error bars indicate detected features, arrows
indicate upper limits. The full line represents the 
line fluxes of \object{MWC 1080} under variable extinction.}
\label{LFhony86.ps}
\end{figure}

Supposing that all HAEBEs have chemically identical PAH molecules, the
PAH spectra of this group of stars should be uniform. Two effects that
could influence the shape of this 'unique' spectrum are the UV field
of the central star and interstellar extinction. If we presume that the
wavelength dependence of the
first does not vary much from source to source, the dominant source of
deviations from a simple linear correlation between the PAH
feature-strengths would be extinction. This will have a stronger effect
on PAH features that are close to the amorphous 10 micron silicate
feature. In Fig.~\ref{LFhony86.ps} the logarithm of the ratio of the
line fluxes of the PAH feature at 8.6 micron over the feature at 6.2
micron, $\log(LF_{8.6}/LF_{6.2})$, is plotted versus the logarithm of
the ratio of the line fluxes of the 3.3 micron PAH feature over the
6.2 feature, $\log(LF_{3.3}/LF_{6.2})$ \citep[following][]{hony01}. 
To include the extinction effect into the plot, we 
applied the interstellar extinction law of \citet{fluks} to the 
line fluxes of \object{MWC 1080} according to the formula
\[  LF'_{j}(x) = LF^{\mathrm{MWC~1080}}_{j} \times 10^{\frac{A_\lambda}{A_V} \frac{x}{2.5}} \] 
in which $x$ represents the actual visual extinction (in mag) and 
$A_\lambda / A_V$ the interstellar extinction coefficient at the
central wavelength 
$\lambda$ of feature $j$. The straight line in Fig.~\ref{LFhony86.ps}
represents the ratios of extinction corrected line fluxes for a wide
range of visual extinction. If interstellar extinction would be the only
effect, all measurements should lie within the observational error bars 
from this line. This is not the case; sources that deviate
significantly are \object{Elias 3-1}, \object{AB Aur},
\object{RR Tau} and \object{CD$-$42$^\circ$11721}. A similar plot based on the PAH
7.7 micron feature instead of the 8.6 micron feature (not in the
paper), leads to the 
same conclusion: the spread cannot be induced by extinction
only. Hence, the PAH spectra of HAEBE stars are not identical,
indicating differences between PAH molecules in different
systems. This spectral variety of PAHs can have several causes. 

\begin{enumerate}
\item As described in the introduction, the different PAH emission
bands studied in this analysis are linked to bending and stretching
modes in the different bonds of the PAH molecules. Differences in PAH
grain sizes will change the ratio of number of CC bonds over number of
CH bonds. Larger grains will have a large CC/CH ratio, which
may lead to e.g. a smaller $LF_{3.3}/LF_{6.2}$ ratio.
\item Differences in PAH ionisation can have a strong effect on the
intensities of the PAH emission features. The PAH charge is ruled by
the photo-ionization rate and the electron recombination rate, and
hence depends strongly on local physical conditions, in particular the
electron density and the strength of the UV radiation field.
\item Chemical differences can also affect the PAH emission
spectrum. PAH molecules can contain other elements than C and H. This
\textit{pollution} influences the strength and especially the peak
position of the feature. The expected shifts in peak position due to
this effect are nonetheless too small to be measurable in the
available spectra. No definite conclusions on the presence of the latter
can be drawn. 
\item Dehydrogenation of PAHs, when the molecules are stripped of
their H atoms, is believed not to play a crucial role
\citep{hony01}. Nevertheless this phenomenon could add to the observed 
differences in the spectra.
\end{enumerate}

The 6.2 micron feature is the most frequently detected PAH feature in our
analysis (25 times out of 46 spectra), followed by the 7.7 micron
feature (19/46). This was to be expected, since these two features are
both linked to the CC bond, which is an indispensable part of the PAH
chemistry. The PAH 3.3 and 8.6 micron features, both linked to the
CH bond, are less frequent (12/45 and 16/46 respectively). 
The mean peak wavelength, FWHM and peak-over-continuum flux ratio of
the examined features is given in Table~\ref{PAHtable}. The mean LF
ratios of all features are summarized in Table~\ref{LFratios}.

\begin{table}
\caption{ The mean peak wavelength $\lambda_0$, FWHM and
peak-over-continuum flux ratio of the detected
features. \label{PAHtable}}
\begin{center}
\begin{tabular}{cccc}
 \hline
 \multicolumn{1}{c}{Feature} &
 \multicolumn{1}{c}{$\langle \lambda_0 \rangle$} &
 \multicolumn{1}{c}{$\langle \mathrm{FWHM} \rangle$} &
 \multicolumn{1}{c}{$\langle \mathrm{PF/CF} \rangle$} \\ 
             & [$\mu$m]&[$\mu$m]&     \\
\hline
PAH 3.3 micron & 3.293 & 0.045 & 1.172 \\
NAN 3.4 micron & 3.415 & 0.067 & 1.284 \\ 
NAN 3.5 micron & 3.530 & 0.059 & 1.542 \\
PAH 6.2 micron & 6.261 & 0.198 & 1.690 \\
PAH 7.7 micron & 7.837 & 0.551 & 1.575 \\
PAH 8.6 micron & 8.641 & 0.311 & 1.310 \\
COMP 11 micron & 11.254& 0.326 & 1.329 \\
Si 9.7 micron  & 9.959 & 2.775 & 1.909 \\
\hline
\end{tabular}
\end{center}
\end{table}

\subsection{The nanodiamond features}

Only a small minority of stars in this sample display nanodiamond
features in their spectrum. Both examined features (at 3.4 and
3.5~$\mu$m) were detected in the spectra 
of \object{Elias 3-1} and \object{HD 97048} \citep[as thoroughly
  described by][]{vankerckhoven}, but also in the spectrum of
\object{BD+40$^\circ$4124}. 
The spectrum of \object{HD 100546} possibly contains the 3.4 micron feature,
while the 3.5 micron feature was detected in the spectrum of \object{MWC 297}
\citep[see also][]{terada}. 

Tables~\ref{PAHtable} and \ref{LFratios} contain general information
about the NAN features. Nevertheless, because of the small number of
sources in this sample that display 3.4 
or 3.5 micron emission, no stringent conclusions on the conditions
needed to see these features can be
drawn. We are currently carrying out a ground-based 3.2--3.6~$\mu$m 
spectroscopic survey of a much larger sample of HAEBEs, which should 
allow us to draw more definitive conclusions on possible correlations 
of the NAN and the PAH features in the near future.

\subsection{The amorphous 10 micron silicate feature \label{the10micron}}

In the sample of HAEBEs presented in this paper, most of the sources
(52\%) have the amorphous 10 micron silicate feature in emission; 6
out of 46 objects (13\%) have the feature in absorption, while it is
undetected in the other cases. We note that at least in some cases, 
the presence of a 10~$\mu$m feature may be masked by the presence 
of strong PAH emission in the 8.6 and 11.2 micron bands, so the 
true fraction of sources with amorphous silicates is likely 
higher than reported here.

\begin{figure}
\rotatebox{0}{\resizebox{3.5in}{!}{\includegraphics{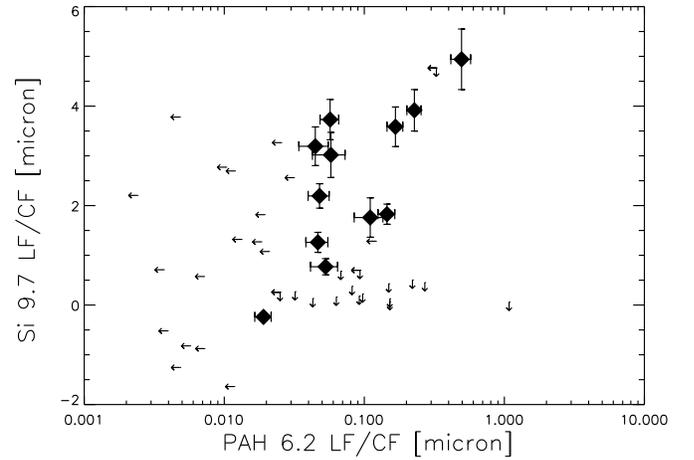}}}
\caption{ The $LF/CF$ ratios of the amorphous 10 micron silicate
feature (Si 9.7) versus the PAH 6.2 micron feature. Diamonds with
error bars indicate detected features, arrows represent upper
limits. Sources with the 10 micron feature in absorption have a
negative $LF/CF$ ratio.}
\label{LFoverCF6297.ps}
\end{figure}

The $LF/CF$ correlation coefficients of the 10 micron feature and the
PAH features are low. In Fig.~\ref{LFoverCF6297.ps} the $LF/CF$ ratio
of the 10 micron feature is plotted versus that of the PAH 6.2 micron
feature. The two ratios correlate poorly, especially when one takes into
account the upper limits (arrows) on the left-hand side in the
plot. These arrows represent spectra in which the 10 micron feature is
present, but where the PAH 6.2 micron feature is undetected. Similar
plots in which the other PAH features are plotted against the 10
micron feature show the same result: the strength of the emission of
PAHs is uncorrelated with the strength of the amorphous 10 micron
silicate feature.

\begin{figure}
\rotatebox{0}{\resizebox{3.5in}{!}{\includegraphics{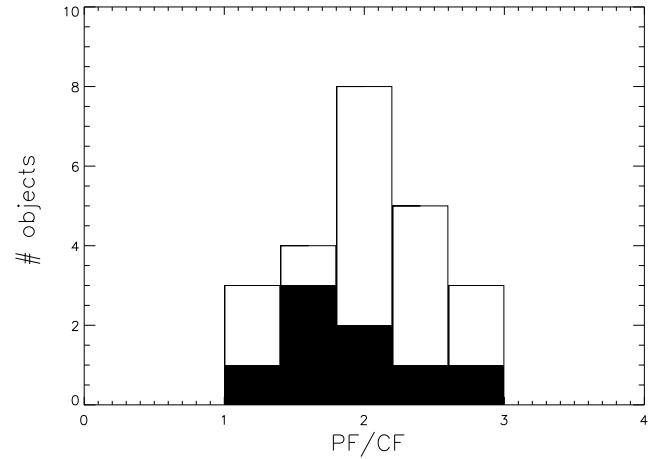}}}
\caption{ The cumulative histogram of the $PF/CF$ for the amorphous 10 micron
silicate feature. The 24 sample stars in which this feature is in emission
are included in this plot. The lower filled part of the bars represents
group I sources, the upper part group II sources.}
\label{PFoverCF97.ps}
\end{figure}

We now consider the peak-over-continuum flux ratio ($PF/CF$)
of the 10 micron feature in our sample. We plotted this
dimensionless quantity for all detected 10 micron emission features in
the sample in Fig.~\ref{PFoverCF97.ps}. The histogram peaks around the
average value 1.91, which means that the peak of the 10 micron
emission is on average twice as high as the underlying continuum. The
minimum and maximum value for the ratio in the sample is 1.21 and 2.69
respectively. 
There does not seem to be a difference between the 10 micron emission
feature in group I and group II sources. The average $PF/CF$
ratios for the two groups are 1.85 and 2.03 respectively. 
Strong as well as more
modest amorphous 10 micron silicate features appear in both group I 
and group II sources. The feature can also be absent in both groups.

\begin{table}
\caption{ The mean line-flux ratios of the examined
features. Indicated in the table is the mean $LF_A/LF_B$ of all sample
stars for which both A and B were detected. \label{LFratios}}
\scriptsize
\begin{center}
\begin{tabular}{@{}l@{}|ccccccc@{}}
\multicolumn{1}{r|}{B}&
           \rotatebox[origin=c]{90}{NAN 3.4} &
           \rotatebox[origin=c]{90}{NAN 3.5} &
           \rotatebox[origin=c]{90}{PAH 6.2} &
           \rotatebox[origin=c]{90}{PAH 7.7} &
           \rotatebox[origin=c]{90}{PAH 8.6} &
           \rotatebox[origin=c]{90}{COMP 11} &
           \rotatebox[origin=c]{90}{Si 9.7} \\ 
\multicolumn{1}{l|}{A} &
                   &         &         &         &         &         &        \\
\hline
PAH 3.3 &  0.640   & 0.310   & 0.171   & 0.074   & 0.271   & 0.138   & 0.010  \\
NAN 3.4 &          & 0.662   & 0.119   & 0.060   & 0.246   & 0.044   & 0.004  \\
NAN 3.5 &          &         & 0.241   & 0.143   & 0.416   & 0.269   & 0.051  \\
PAH 6.2 &          &         &         & 0.437   & 1.430   & 0.802   & 0.051  \\
PAH 7.7 &          &         &         &         & 3.302   & 1.830   & 0.113  \\
PAH 8.6 &          &         &         &         &         & 0.443   & 0.026  \\
COMP 11 &          &         &         &         &         &         & 0.043  \\
\end{tabular}
\end{center}
\normalsize
\end{table}

\subsection{The 11 micron feature}

\begin{figure}
\rotatebox{0}{\resizebox{3.5in}{!}{\includegraphics{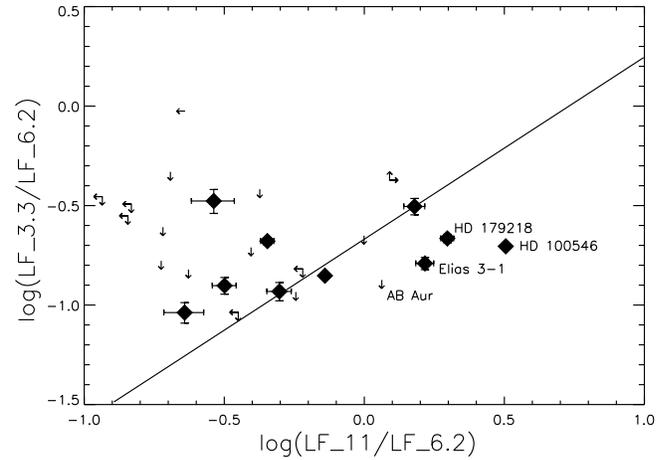}}}
\caption{ Same type of plot as Fig.~\ref{LFhony86.ps}. 
The logarithm of the ratio of the line fluxes of the PAH
features at 3.3 and 6.2 micron $\log(LF_{3.3}/LF_{6.2})$ versus the
logarithm of the ratio of the line fluxes of the features at 11 and
6.2 micron $\log(LF_{11}/LF_{6.2})$. Again, the
full line represents the line fluxes of \object{MWC 1080} under variable
interstellar extinction.}
\label{LFhony11.ps}
\end{figure}

The 11 micron feature is detected in 48\% of the spectra. It is
unclear how many of these detected features include the PAH 11.2
micron band, since this feature can be blended with the crystalline
silicate feature at 11.3 micron. Fig.~\ref{LFhony11.ps}
is similar to Fig.~\ref{LFhony86.ps}. The logarithms of the
ratios $LF_{11}/LF_{6.2}$ and $LF_{3.3}/LF_{6.2}$ are plotted. Again,
the straight line indicates the variable extinction line, based on the
line fluxes of \object{MWC 1080}. Four sources (\object{AB Aur},
\object{Elias 3-1}, \object{HD 100546} and \object{HD 179218}) lie
significantly to the right of 
this line. This suggests that these sources have an 11 micron
feature that contains both PAH \textit{and} crystalline silicate
emission. The presence of crystalline silicates in \object{HD 100546} and 
\object{HD 179218} was already well-established from longer-wavelength 
ISO data \citep{malfait, malfait99}. The detection of 
crystalline silicates in \object{AB Aur} and \object{Elias 3-1} is a new result 
from our quantitative analysis.  
We stress that the detection of a stronger than expected 
11 micron feature in the four sources mentioned above 
does not mean that other sources cannot have crystalline 
silicates; it could just be masked by the presence 
of strong PAH emission at 11.2~$\mu$m, or the crystalline 
silicates could be cooler than a few hundred K, causing 
only very weak emission in our diagnostic band at 11 microns.

\begin{figure}
\rotatebox{0}{\resizebox{3.5in}{!}{\includegraphics{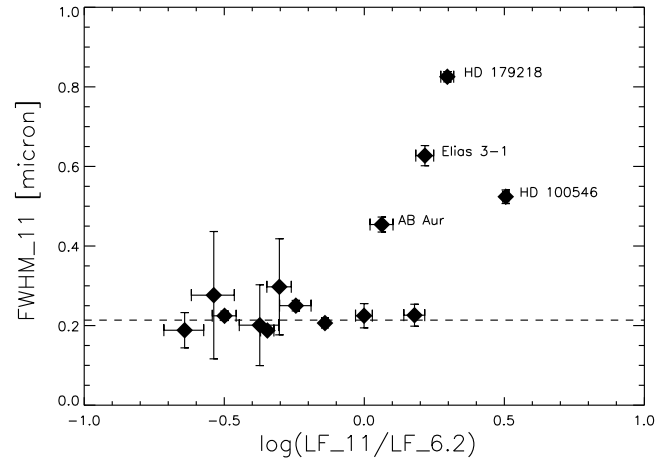}}}
\caption{ The FWHM of the 11 micron feature versus the logarithm of
the ratio of the line fluxes of the features at 11 and 6.2 micron
$\log(LF_{3.3}/LF_{6.2})$. The dashed line represents the average FWHM
(0.214 micron) when excluding the 4 outliers.}
\label{FWHMvsLF11.ps}
\end{figure}

\begin{figure}
\rotatebox{0}{\resizebox{3.5in}{!}{\includegraphics{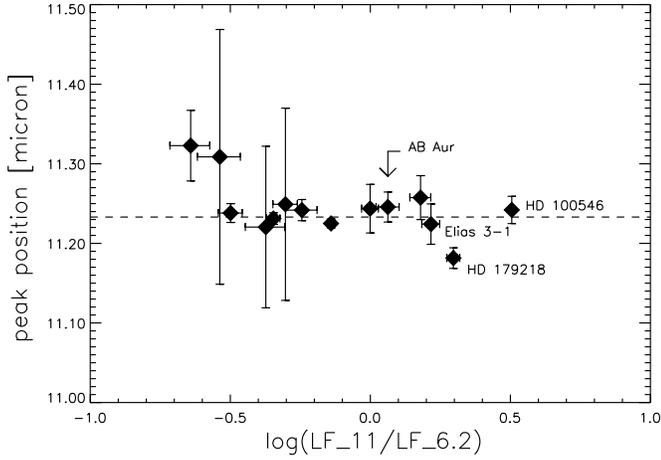}}}
\caption{ The peak position of the detected 11 micron features versus
the ratio of the line fluxes of the 11 micron feature over the 6.2 PAH
feature $\log(LF_{11}/LF_{6.2})$. The dashed line at 11.233 micron
indicates the mean peak position of the 11 micron feature.}
\label{PWhony11.ps}
\end{figure}

When plotting the FWHM of the feature against the ratio
$\log(LF_{11}/LF_{6.2})$ (Fig.~\ref{FWHMvsLF11.ps}), the same four
sources appear to have much higher FWHMs. The horizontal line in the plot
indicates the average FWHM (with exclusion of the 4 objects mentioned
before). This also points to the presence of 2 blended emission
features. Even so, the peak position of the 11 micron feature in \object{AB
Aur}, \object{Elias 3-1}, \object{HD 100546} and \object{HD 179218} is comparable with the average
peak wavelength (see Fig.~\ref{PWhony11.ps}). This might indicate that
the peak position of the crystalline silicate feature varies, which
can be attributed to a variety in its mineorological composition. We stress
however that the uncertainties in these determinations are large.

Sample stars that do not display significant PAH emission at 3.3, 6.2,
7.7 or 8.6 micron are not expected to have significant PAH 11.2 micron
emission. Hence a detected 11 micron feature in the spectra of these sources
is likely to be solely due to crystalline silicate emission. In the
present sample \object{HD 142527} (group I), \object{HD 142666}, \object{HD
144432}, \object{HR 5999}, \object{HD 150193}, \object{HD 163296} and
\object{VV Ser} (all group II) meet these requirements. If these objects
indeed display the crystalline silicate band in their spectrum, then
the number of sources with crystalline silicates in this sample is \textit{at
least} 11 (24\%).

The 11 micron complex does not correlate with the PAH 6.2 micron
feature, and reaches better agreement with other CH bond related
PAHs (3.3, 7.7 micron), which is expected, since the 11.2 micron 
feature is due to a CH bond itself. The correlation with
the PAH 8.6 micron feature, on the other hand, is weak.

\begin{figure}
\rotatebox{0}{\resizebox{3.5in}{!}{\includegraphics{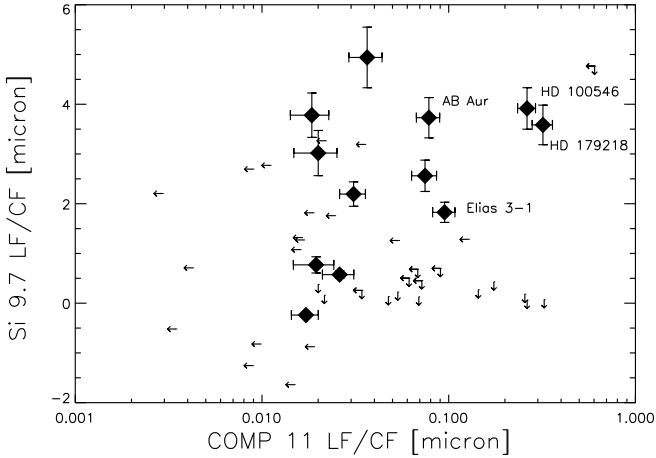}}}
\caption{ The $LF/CF$ ratios of the amorphous 10 micron silicate
feature (Si 9.7) versus the 11 micron feature. Diamonds with error
bars indicate detected features, arrows represent upper
limits. Sources with the 10 micron feature in absorption have a
negative $LF/CF$ ratio.}
\label{LFoverCF1197.ps}
\end{figure}

In Fig.~\ref{LFoverCF1197.ps} the ratio $LF/CF$ of the amorphous
silicate feature versus the 11 micron complex is plotted. There does
not seem to be a correlation between the two features. This can be 
interpreted in two ways. The 11 micron feature
could be mostly dominated by PAH emission in this sample of stars,
which leads to a poor correlation with the 10 micron feature like for the
other PAH features (see Sect.~\ref{the10micron}). The other
possibility is that the crystalline 11.3 micron silicate band
---when present in the 11 micron complex--- does not correlate with the
amorphous 10 micron silicate feature. Note that also the four
'crystalline silicate stars' mentioned above do not display an obvious
trent. This could be interpreted in terms of the second explanation.

\begin{figure}
\rotatebox{0}{\resizebox{3.5in}{!}{\includegraphics{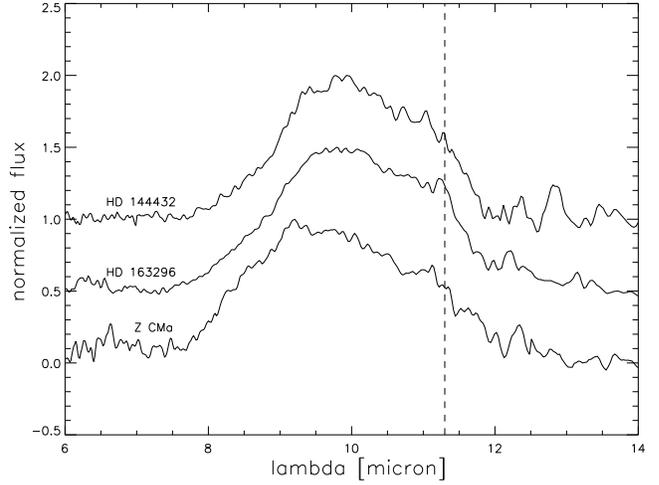}}}
\caption{ The inverted spectrum of \object{Z CMa}, and the emission band
spectra of \object{HD 144432} and \object{HD 163296}. For \object{Z CMa},
the figure shows the inverted continuum-divided, continuum-subtracted
flux $[\mathcal{F}_{cont}(\lambda) - \mathcal{F}(\lambda)]/
\mathcal{F}_{cont}(\lambda)$. The other two spectra display the
continuum-subtracted flux. All three spectra are normalized to a peak
flux of 1. An additional offset was applied for the sake of
clarity. The dashed line represents $\lambda = 11.3~\mu$m}
\label{ZCMAcryst.ps}
\end{figure}

Except for \object{Z CMa}, no crystalline silicate absorption or
emission is observed in sources which display amorphous silicate
absorption (group III). The spectrum of \object{Z CMa} does display a
local minimum around 10.7--11.5 micron, which might be due
to crystalline silicate absorption. In Fig.~\ref{ZCMAcryst.ps}, the
spectrum of \object{Z CMa} is plotted. For reasons of clearness it was
inverted, making the absorption features appear as emission
features. Two sources (\object{HD 144432} and \object{HD 163296}) that are
likely to display a crystalline silicate emission feature at 11.3
micron, are also plotted as a reference.

\section{Correlations between solid-state features and disk properties}

\subsection{The SED}

\begin{figure}
\rotatebox{0}{\resizebox{3.5in}{!}{\includegraphics{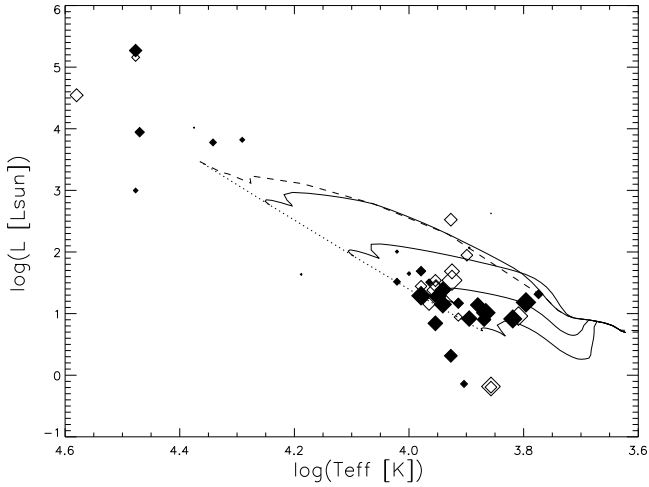}}}
\caption{ The Hertzsprung-Russel diagram for the sample stars. The
plotting symbols are proportional to the PAH luminosity. Filled
diamonds indicate detected emission, open symbols represent upper
limits. The full lines indicate the PMS evolutionary tracks of 
stars with masses M = 1.5, 2, 3 and 5 M$_{\odot}$ from bottom to
top; the dashed and dotted lines represent respectively the ZAMS and
the MS \citep{bernasconi}.}
\label{HRDLPAH.ps}
\end{figure}

\begin{figure}
\rotatebox{0}{\resizebox{3.5in}{!}{\includegraphics{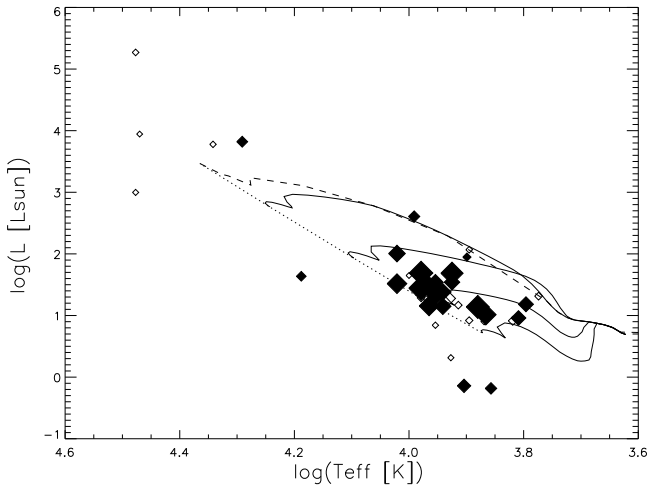}}}
\caption{ Similar plot as Fig.~\ref{HRDLPAH.ps}. The plotting symbols are
proportional to the peak-over-continuum flux of the amorphous 10
micron silicate emission feature.}
\label{HRDPFSi.ps}
\end{figure}

Using the luminosities and effective temperatures computed in 
Sect.~\ref{thesed}, we plot the stars in our sample in a traditional
Hertzsprung-Russel (HR) diagram. There is no
strong correlation between the classification of the sources and the
position in the HR diagram. In Fig.~\ref{HRDLPAH.ps}, the plotting
symbols are proportional to the observed PAH luminosity ($L_{\mathrm{PAH}}$; 
Appendix). In this plot, theoretical
pre-main-sequence (PMS) evolutionary tracks of stars with masses M = 1.5, 2, 3
and 5 M$_{\odot}$ are plotted \citep{bernasconi}. Furthermore,
the zero-age main sequence (ZAMS) is indicated by a dashed line. Most
of the stars lie close to the main sequence (MS), which is represented by the
dotted line. Based on this diagram, the masses of the bulk of the sample
stars lie between 1.5 and 3 M$_{\odot}$. 
The strength of the PAH bands does not seem to be strongly correlated
with stellar mass or age. Fig.~\ref{HRDPFSi.ps} is a plot similar to
Fig.~\ref{HRDLPAH.ps} for the amorphous 10 micron silicate emission
feature. The plotting symbols are proportional to the
peak-over-continuum flux PF/LF of the feature. The strength of
this solid-state band appears to be also uncorrelated with stellar mass and
age.

One of the main characteristics of HAEBE stars is the IR excess due to
thermal emission of circumstellar dust. In Table~\ref{parameters}, the ratio of this
excess luminosity $L_{\exc}$ over the stellar luminosity $L$ is
included. Ten of the sample sources have an IR-excess luminosity that
is larger than the stellar luminosity. This means that these
sources cannot have passive reprocessing circumstellar disks.
There does not seem to be a link between the classification of the
sample stars and their $L_{\exc}/L$ ratio. The average value for the
sources for which the ratio is smaller than unity, is $0.37 \pm 0.21$
for group I, $0.39 \pm 0.20$ for group II and $0.09 \pm 0.04$ for group III.
Half of the group III objects have a ratio bigger than one, though, which is
consistent with the idea that these sources are still actively
accreting systems.

\begin{figure}
\rotatebox{0}{\resizebox{3.5in}{!}{\includegraphics{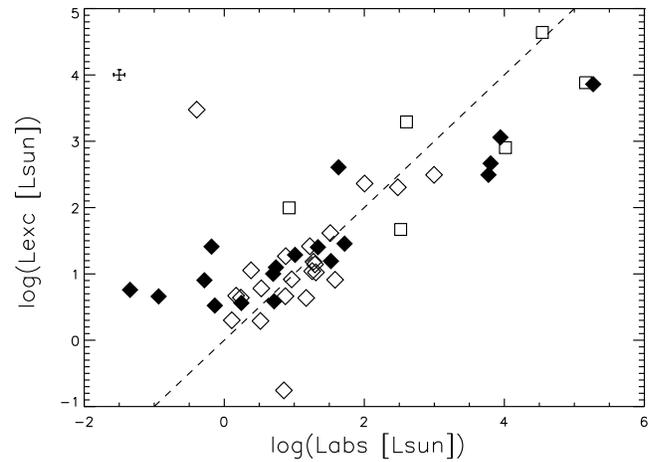}}}
\caption{ The logarithm of the IR-excess luminosity $\log(L_{\exc})$ versus the
logarithm of the absorbed luminosity $\log(L_{\abs})$. Filled diamonds
represent group I, open diamonds group II and squares group III
sources. The dashed line is the first bisector. In the left upper
corner, the typical error bars are represented by a cross.}
\label{LabsvsLexc.ps}
\end{figure}

In Fig.~\ref{LabsvsLexc.ps} the logarithm of the IR-excess luminosity
$\log(L_{\exc})$ is plotted versus the logarithm of the absorbed
luminosity $\log(L_{\abs})$. The dashed line represents $L_{\exc} =
L_{\abs}$. Most of the sample stars lie close to this line. 
Exceptions include the group II source in the upper left corner which
represents \object{R CrA} and the open diamond at the bottom of the plot,
representing \object{HD 141569}. The IR excess luminosity of \object{R CrA} is
due to confusion with the nearby source \object{CrA IRS7} \citep{choi}.
Furthermore, a set of group I sources with low $L_{\abs}$
values, display relatively high $L_{\exc}$ values. An explanation for this
\textit{extra} IR excess luminosity could be the presence of a
late-type companion. Also the orientation of the system can play a role,
especially in group I sources, where the flaring of the disk implies that the
line of sight to the central star already passes through the disk at
fairly low inclinations (i $\sim$ 45$^\circ$).

\subsection{The SED versus the spectral features}

The different parameters that characterize the SED were compared
to the parameters that describe the infrared solid state bands. In
this section we summarize the results.

\begin{figure}
\rotatebox{0}{\resizebox{3.5in}{!}{\includegraphics{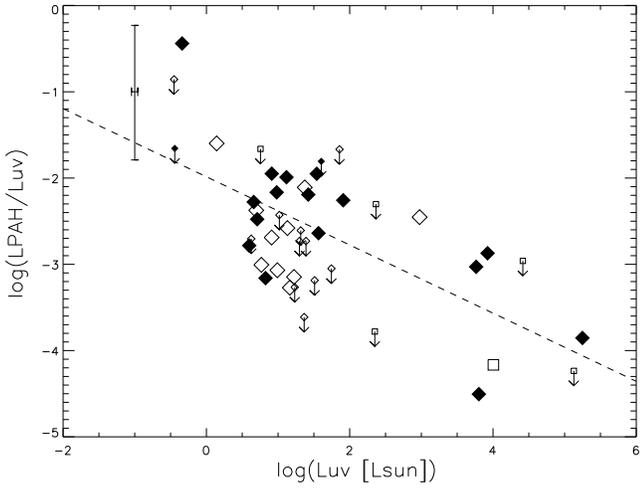}}}
\caption{ The PAH-over-UV luminosity $\log(L_{\mathrm{PAH}}/L_{\uv})$ versus the
intensity of the UV radiation field $\log(L_{\uv})$. The PAH luminosity
is the sum of the luminosities of the PAH features at 3.3, 6.2, 7.7
and 8.6 micron. In the left upper corner, the cross indicates the
typical errors. Filled diamonds refer to
group I, open diamonds to group II and squares to
group III objects. Arrows represent upper limits. The dashed line
represents the best linear fit to the data:
$\log(L_{\mathrm{PAH}}/L_{\uv}) = -0.395 \log(L_{\uv}) - 1.99$.}
\label{LuvLFdall.ps}
\end{figure}

The UV luminosity of the central star correlates with
the PAH luminosities: the absolute PAH luminosity increases with
increasing UV radiation. The ratio $L_{\mathrm{PAH}}/L_{\uv}$ however
\textit{decreases} with increasing UV luminosity. In
Fig.~\ref{LuvLFdall.ps}, the latter correlation is demonstrated. The
dashed line in this figure represents $\log(L_{\mathrm{PAH}}/L_{\uv})
= -0.395 \log(L_{\uv}) - 1.99$ or equivalently $L_{\mathrm{PAH}} =
0.01~L_{\uv}^{0.6}$. If the emitted PAH luminosity would
increase linearly with increasing UV radiation of the central star
($\log(L_{\mathrm{PAH}} \propto L_{\uv}$), the ratio
$L_{\mathrm{PAH}}/L_{\uv}$ would have been constant for all
$L_{\uv}$. This is obviously not the case. A higher stellar UV
luminosity indeed increases the PAH luminosity, but does this in a
\textit{sub-linear} way: the PAH luminosity increases with $L_{\uv}^p$
where $p=0.6<1$. It appears that either the
efficiency of the 
absorption/emission process decreases with increasing UV strength, or
that the increasing \textit{hardness} of the UV photons plays a role.
Since hotter stars emit not only more UV photons, but also relatively
more high-energy UV photons, the observed trend might be a contrast effect.
Short-wavelength UV photons cannot be
absorbed by PAHs (no increase of $L_{\mathrm{PAH}}$), but do
contribute to the total UV luminosity $L_{\uv}$, hence decreasing
the luminosity ratio.

It is quite remarkable that some group II sources, even though they have
comparable UV luminosities as their group I counterparts, do not
display PAH emission. This indicates that the appearance of PAH emission
is linked to the group I/II classification of the sources, and does not 
depend on $L_{\uv}$ only.

Suppose that the IR emission is due to PAH molecules that
are homogeneously distributed in a spherical, optically thin halo around the star,
which has the radius of the ISO-SWS beam (11'') at the appropriate
distance. It is then possible to estimate the mean particle density of the PAH
molecules ($\rho$; see Appendix) in that halo. In Table~\ref{NISO} the computed values
are listed. The detected emission of the PAH features at 3.3, 6.2,
7.7 and 8.6 micron has been included. Note that not all emitted PAH
flux (e.g. the PAH 11.2 micron flux) is accounted for. This implies
that ---under the given assumptions--- the computed molecule density is a
lower limit.
For most objects, the resulting values are high compared to the
typical ISM PAH density ($\rho_{\mathrm{typ}} = 5 \times 10^{-7}~\cm^{-3}$;
Appendix). We interpret this as an indication that the PAH emission
does not emanate from a halo around the central star. For most stars, 
a denser environment is needed to explain the observed emission.

\begin{table}
\caption{ Supposing a homogeneous distribution of the PAH molecules in
a halo around the central star, with a radius of the size of the
ISO-SWS beam (11'') at the appropriate distance, the PAH particle density $\rho$ is
computed based on the $L_{\mathrm{PAH}}/L_{\uv}$ ratio. All objects in which PAH
emission is detected are adopted in this table and are listed
according to decreasing PAH-over-UV luminosity.
 \label{NISO}}	  
\begin{center}
\begin{tabular}{|c|c|cc|}
 \hline
 \multicolumn{4}{|c|}{\bf Does PAH emission emanate from a halo?}\\ \hline \hline
 \multicolumn{1}{|c|}{Object} &
 \multicolumn{1}{|c|}{Group} &
 \multicolumn{1}{|c}{$L_{\mathrm{PAH}}/L_{\uv}$} &
 \multicolumn{1}{c|}{$\rho$} \\ 
                         &     &               & [$\cm^{-3}$]     \\
\hline \hline
 \object{Elias 3-1}               & I   &    0.36       & 5.51 10$^{-4}$   \\
 \object{RR Tau}                  & II  & 2.5 10$^{-2}$ & 3.82 10$^{-5}$   \\
 \object{HD 34700}                & I   & 1.1 10$^{-2}$ & 8.13 10$^{-5}$   \\
 \object{HD 97048}                & I   & 1.1 10$^{-2}$ & 1.51 10$^{-5}$   \\
 \object{HD 34282}                & I   & 1.0 10$^{-2}$ & 6.21 10$^{-6}$   \\
 \object{VX Cas}                  & II  & 7.7 10$^{-3}$ & 3.00 10$^{-6}$   \\
 \object{HD 169142}               & I   & 6.8 10$^{-3}$ & 1.14 10$^{-5}$   \\
 \object{HD 100546}               & I   & 6.4 10$^{-3}$ & 1.51 10$^{-5}$   \\
 \object{HD 179218}               & I   & 5.5 10$^{-3}$ & 5.60 10$^{-6}$   \\
 \object{HD 100453}               & I   & 5.2 10$^{-3}$ & 1.14 10$^{-5}$   \\
 \object{HD 95881}                & II  & 4.2 10$^{-3}$ & 8.70 10$^{-6}$   \\
 \object{Wra 15-1484}             & II  & 3.5 10$^{-3}$ & 1.14 10$^{-6}$   \\
 \object{HD 139614}               & I   & 3.3 10$^{-3}$ & 5.78 10$^{-6}$   \\
 \object{VV Ser}                  & II  & 2.6 10$^{-3}$ & 1.93 10$^{-6}$   \\
 \object{AB Aur}                  & I   & 2.2 10$^{-3}$ & 3.87 10$^{-6}$   \\
 \object{HD 142666}               & II  & 2.0 10$^{-3}$ & 3.42 10$^{-6}$   \\
 \object{HD 135344}               & I   & 1.6 10$^{-3}$ & 2.86 10$^{-6}$   \\
 \object{CD$-$42$^\circ$11721}    & I   & 1.3 10$^{-3}$ & 8.16 10$^{-7}$   \\
 \object{HD 144432}               & II  & 9.8 10$^{-4}$ & 1.65 10$^{-6}$   \\
 \object{BD+40$^\circ$4124}       & I?  & 9.3 10$^{-4}$ & 2.31 10$^{-7}$   \\
 \object{HD 31648}                & II  & 8.5 10$^{-4}$ & 1.58 10$^{-6}$   \\
 \object{HD 163296}               & II  & 7.1 10$^{-4}$ & 1.42 10$^{-6}$   \\
 \object{HD 142527}               & I   & 6.9 10$^{-4}$ & 1.15 10$^{-6}$   \\
 \object{HD 141569}               & II  & 5.3 10$^{-4}$ & 1.31 10$^{-6}$   \\
 \object{MWC 1080}                & I   & 1.4 10$^{-4}$ & 1.54 10$^{-8}$   \\
 \object{MWC 297}                 & III & 6.8 10$^{-5}$ & 6.64 10$^{-8}$   \\
 \object{HD 200775}               & I   & 3.1 10$^{-5}$ & 1.72 10$^{-8}$   \\
 \hline			  
\end{tabular}		  
\end{center}		  
\end{table}

\begin{figure}
\rotatebox{0}{\resizebox{3.5in}{!}{\includegraphics{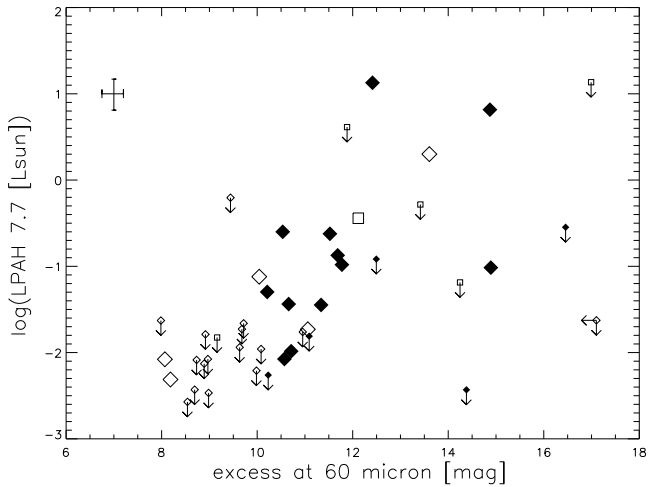}}}
\caption{ The PAH 7.7 micron luminosity $log(L_{\mathrm{PAH 7.7}})$ versus the
excess at 60 micron (in magnitudes). 
Filled diamonds represent group I, open diamonds
group II and squares group III sources. Small symbols indicate undetected
features, arrows stand for upper limits. In the left upper corner the
typical error bars are displayed.}
\label{exc60vs77.ps}
\end{figure}

\citet{habart} have modelled PAH emission in flared circumstellar
disks. In their models, they assume PAH molecules containing
40--100 carbon atoms. For PAHs of  
these sizes, the influence of ionization and photo-evaporation affects
the line-flux ratios of the PAH features, in particular the relative
strength of the 3.3 micron feature. Even though a clear dependence of the
PAH line fluxes on the stellar UV luminosity was observed in our
present sample, no 
correlations between the \textit{relative} strength of the PAH 
features and the UV radiation were noted. The
line-flux ratio of the 3.3 and 6.2 micron feature --which
ranges from 9 to 94\%-- is independent of the central star's UV
field.

The excesses in the K band, at 60, 850 and 1300 micron were compared to
the PAH data. Fig.~\ref{exc60vs77.ps} shows the luminosity of the PAH
7.7 feature 
versus the excess at 60 micron. The sources with the faintest 60
micron excesses (by definition mostly group II objects) are those that
have the faintest PAH emission. Other PAH features display comparable
behaviour.

Notwithstanding these correlations, in general the luminosities of the IR emission
bands do not correlate well with the other measured
excesses. The lack of correlation between the 850 or 1300 micron
excess and the strength of the PAH features shows that 
the disk mass has no influence on the resulting PAH spectrum. This is
consistent with the idea that PAH emission emanates from
the surface layers (\textit{atmosphere}) of the circumstellar 
\textit{disk}, which contains only a small fraction of the disk
mass. The FWHMs of the IR PAH bands do not correlate with the
excesses either. 

The Rayleigh-Jeans part of the SED can be modelled with a power law,
$\log(\lambda \mathcal{F}_\lambda) \sim \lambda^{-n}$ in which n is the
\hbox{(sub-)mm} spectral index. There is no clear correlation between this
index and the strength of  
the PAH features. The sub-mm slope of the spectrum can be a proxy for
the grain size 
distribution of the cold, large grains in the outer parts and
mid-plane of the disk \citep[][henceforth A04]{ackesubmm}. This
non-correlation indicates that the grain 
size distribution of the bulk of the disk's material has no influence
on the PAH emission. This is again in agreement with the hypothesis that PAH
molecules are excited in and radiate from the disk's atmosphere.

The parameters that describe the amorphous silicate 10 micron feature do not
correlate with the IR flux excesses. This indicates that the strength
of the 10 micron feature does not depend on the disk mass. 
The LF and FWHM of the silicate band do not correlate with the sub-mm spectral
index either. 
The grain size distribution of
the cold particles in the outer parts of the disk and the warm small
silicate grains are independent of each other. Nevertheless, if
vertical mixing in the disk is efficient, one may expect that only the
smallest grains make it to the surface. Hence the amorphous 10 micron
silicate feature may be indicative for the smallest grain sizes
(still) present in the disk (A04).

We compared the SEDs of the four stars for which the 11 micron feature
is a blend of the PAH 
11.2 micron band and the crystalline silicate feature at 11.3 micron
(\object{Elias 3-1}, \object{AB Aur}, \object{HD 100546} and \object{HD
179218}) with the SEDs of the other sample 
stars. The sub-mm spectral index $n$, 10 micron $PF/CF$ ratio and 1300
micron flux excess of these group I sources make them normal
members of this group. The objects \object{AB Aur}, \object{HD 100546} and
\object{HD 179218} have steep sub-mm slopes (n$\sim$4.20), high 10
micron $PF/CF$ values (larger than the average value 1.91), and
intermediate 1300 micron excesses ($\sim$10 mag). 
The circumstellar
material of these sources thus contains cold grains in the outer parts
of the disk, significantly smaller than mm-sized (AV04) and small
($\sim$0.1~$\mu$m) warm silicate grains \citep{bouwman01, vanboekel}. 
The estimated disk mass is of the order of 1\% of the 
stellar mass. \object{Elias 3-1} on the other hand is somewhat more
peculiar, since $n$ is equal to 3.22 and the 10 micron peak-over-continuum flux
ratio is 1.70. The circumstellar matter in this system
seems to be more evolved. The disk mass of this source is also high
compared to the stellar mass (30\%). This is close to the theoretical
upper limit for gravitational stability \citep[e.g.][]{gammie}.


\section{Interpretation}

\begin{figure}
\rotatebox{0}{\resizebox{3.5in}{!}{\includegraphics{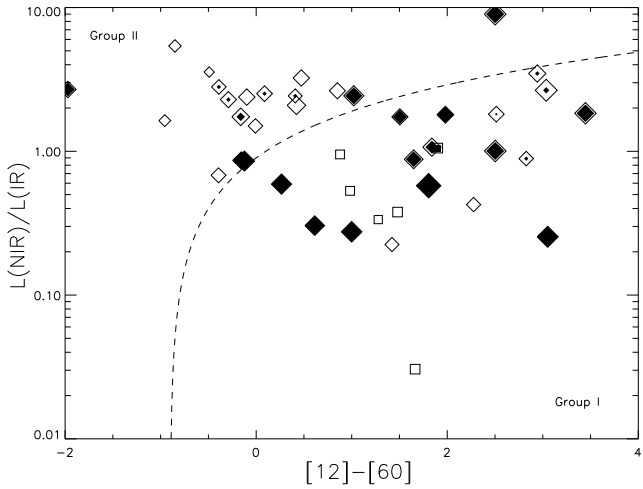}}}
\caption{ Similar plot as Fig.~\ref{plotroy.ps}. The plotting symbols
are scaled proportional to the strength ($\sum LF/CF$) of the PAH features
at 3.3, 6.2, 7.7 and 8.6 micron. The outer limit of the
partially filled plotting symbol represents the upper limit for the strength;
the filled inner part refers to the part that was actually
detected. Diamonds represent group I and group II sources, squares
group III objects.}
\label{plotroyall.ps}
\end{figure}

We use plotting diagrams like Fig.~\ref{plotroy.ps}, and include the
strength of the IR spectral features by scaling the plotting symbol
proportional to $LF/CF$.
In Fig.~\ref{plotroyall.ps}, the sum of the $LF/CF$ ratios of the PAH
features at 3.3, 6.2, 7.7 and 8.6 micron is
indicated in this manner. The filled diamonds in the plot represent
detected features, while open diamonds stand for upper limits. Squares
refer to group III sources. The sizes of the diamonds and squares represent
the strength of the PAH features on a logarithmic
scale. We specifically indicate group III sources since, because of
their location in 
the diagram, they could be confused with group I sources.
That group III sources appear in the \textit{group I part} of the plot is
to be expected, since these sources are believed to 
be in an earlier evolutionary stage; the objects are still highly
embedded in the circumstellar environment, which blocks
out the stellar light. This makes their SEDs very red and
hence they show up in the lower right part of the diagram. 

It is striking that the sources that display the strongest PAH
emission are mostly group I sources. The \citet{dullemond02} model
suggests that these objects  
have circumstellar disks with a flaring geometry. PAH molecules in the
atmosphere of the flaring part of these disks are directly irradiated
by the UV photons of the central star. This implies that the molecules
will be excited and will radiate in the IR, no matter how far out in
the disk, as long as the UV field is strong enough.
PAH molecules in the atmosphere of the self-shadowed, geometrically flatter
disks of group II sources barely see any direct UV radiation, because
the puffed-up inner rim blocks out most of the starlight. Hence, in these
systems the IR emission bands are expected to be much fainter. 
Most objects in our sample are consistent with this hypothesis.

Except for \object{MWC 297}, no PAH emission is detected in group III
sources. The circumstellar matter of these highly embedded objects can
probably accurately shield the stellar UV radiation close to the
central star. The volume of excited PAH molecules will likely be
too small to be observable. Furthermore, the PAH IR radiation has
to travel through the dense circumstellar environment. This prohibits
the appearance of PAH emission in the group III spectra.

\begin{figure}
\rotatebox{0}{\resizebox{3.5in}{!}{\includegraphics{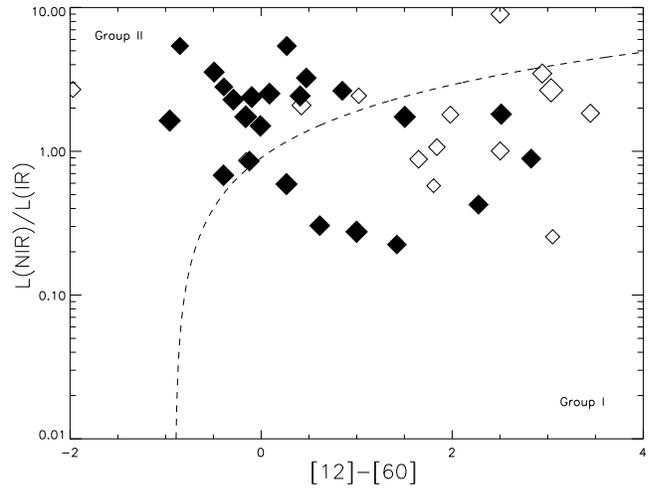}}}
\caption{ Similar plot as Fig.~\ref{plotroyall.ps} for the amorphous 10
micron silicate feature. The plotting symbols are scaled proportional to the
$LF/CF$ ratio of the feature. The filled diamonds indicate stars with a
detected 10 micron emission feature, the open diamonds indicate
upper limits.}
\label{plotroy97.ps}
\end{figure}

In Fig.~\ref{plotroy97.ps} the $LF/CF$ ratio of the amorphous 10
micron silicate feature is plotted in the
$L_{\mathrm{NIR}}/L_{\mathrm{IR}}$ versus 
$[12]-[60]$ diagram. Again, the filled diamonds refer to the detected
emission features, while the open diamonds represent upper
limits. Group III sources, which display amorphous silicate \textit{absorption},
are not included in this plot. Note that the
silicate emission feature appears in both groups. When present, the
strength of the emission feature seems to be independent of the
classification of the source. Nevertheless, a larger fraction of group
I sources have an undetected amorphous silicate feature (10/18 versus 6/22 in
group II). M01 suggested that this is probably a selection effect; group II
objects that do have 10 micron silicate emission are brighter, and
hence are observationally favoured over group II sources without
silicate emission. This selection effect plays a smaller role for
group I sources, since they are by definition brighter IR emitters.

\begin{figure}
\rotatebox{0}{\resizebox{3.5in}{!}{\includegraphics{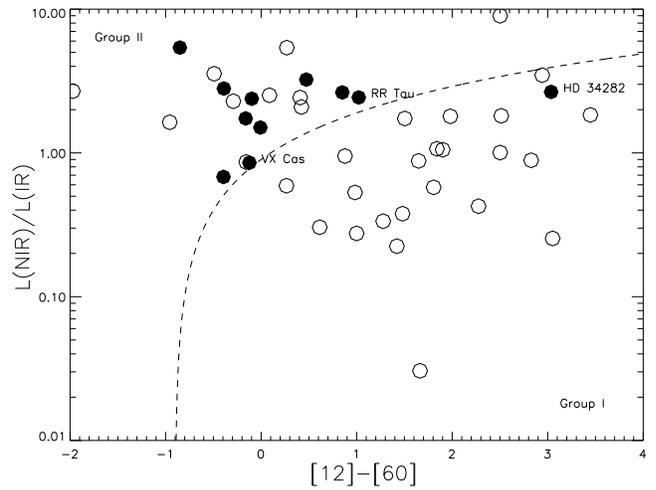}}}
\caption{ Similar plot as Fig~\ref{plotroy.ps}. The filled symbols
indicate UXORs, the open symbols are the other HAEBEs in this
sample. The 3 UXOR stars that display clear PAH emission are \object{VX Cas},
\object{HD 34282} and \object{RR Tau}. }
\label{plotroyuxors.ps}
\end{figure}

According to the definition of \citet{dullemond03}, 13 of our
sample sources are UX Orionis (UXOR) stars. In their paper, the
authors argue that the disk geometry of this class of stars must be
self-shadowed and hence that UXORs are group II objects. In the
present paper we have shown that PAH emission in group II sources is
weak. Of the 13 UXORs in our sample, only 3 sources
\textit{do show} significant PAH emission: \object{VX Cas}, \object{HD 34282} 
and \object{RR Tau}. They are labelled in
Fig.~\ref{plotroyuxors.ps}. We note that all three lie close to the 
empirical separation line between group I and II; they might be 
transitional objects in which the shadow cast by the inner rim is 
relatively small.  We remind the reader that
\object{R CrA} and \object{LkH$\alpha$ 224} are not in the plot, and that
these sources were classified \textit{based on} their UXOR behaviour.

\begin{figure}[!h]
\rotatebox{0}{\resizebox{3.in}{!}{\includegraphics{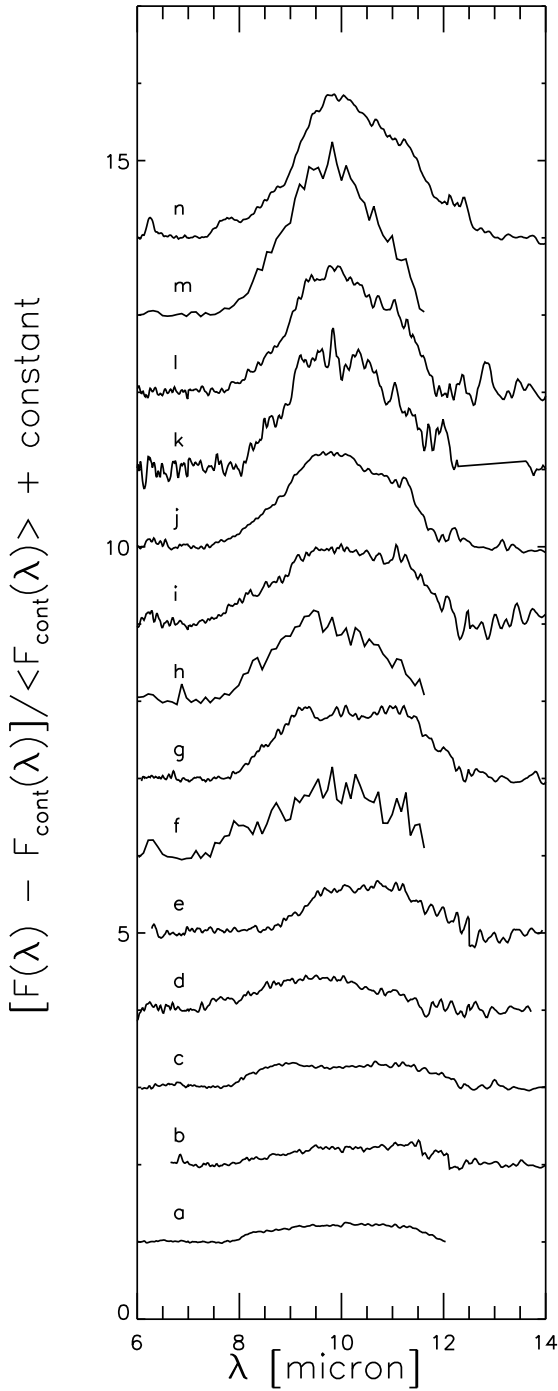}}}
\caption{The amorphous 10 micron silicate band for 14 of the sample
stars. The figure shows the continuum-subtracted flux
$\mathcal{F}(\lambda) - \mathcal{F}_{\mathrm{cont}}(\lambda)$, which was
divided by the average value of the underlying continuum
$\langle\mathcal{F}_{\mathrm{cont}}(\lambda)\rangle$. The spectra of
the stars are sorted 
by increasing peak-over-continuum flux from a to n: a--\object{R CrA}; b--\object{HR
5999}; c--\object{V376 Cas}; d--\object{HD 200775}; e--\object{T CrA};
f--\object{VX Cas}; g--\object{HD 104237}; 
h--\object{BF Ori}; i--\object{HD 31648}; j--\object{HD 163296};
k--\object{SV Cep}; l--\object{HD 144432}; m--\object{UX Ori};
n--\object{AB Aur}. For clarity, the zero levels of the
14 spectra are shifted.}
\label{Siemission.ps}
\end{figure}

\begin{figure}
\rotatebox{0}{\resizebox{3.in}{!}{\includegraphics{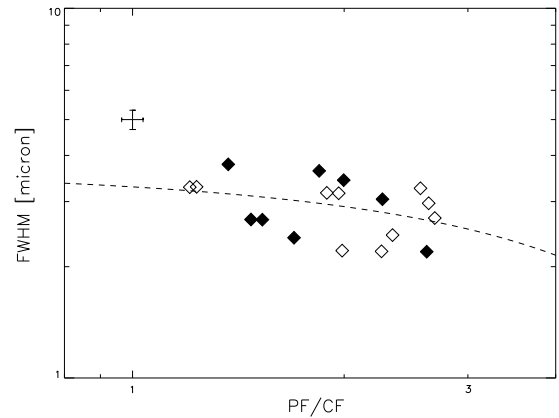}}}
\caption{ The FWHM of the amorphous 10 micron silicate feature versus
the peak-over-continuum flux ratio $PF/CF$. Filled and open diamonds
refer to group I and group II sources respectively. The dashed line
represents the best fit to the data; $\mathrm{FWHM} = -0.38 \times
\mathrm{PF/CF} + 3.67$. The typical error bar is indicated in the
upper left corner.}
\label{FWHMvsPF97.ps}
\end{figure}

In Fig.~\ref{Siemission.ps}, a selection of 14 of our
sample stars shows that the 10 micron emission feature tends to
become broader when the peak-over-continuum flux ratio
decreases. Small warm silicate grains ($\sim$0.1 $\mu$m) cause a
distinct feature that 
peaks around 9.7 micron, while larger grains ($\sim$2 $\mu$m) induce a
broader, less pronounced feature with a peak towards longer
wavelengths \citep{bouwman01}. \citet{vanboekel} interpret this
sequence in terms of grain growth in the disk. In
Fig.~\ref{Siemission.ps} this sequence is reproduced. Our analysis
confirms observationally the correlation between the shape and the strength of the
amorphous 10 micron silicate band. In their \textit{Letter},
\citet{vanboekel} only plotted group II sources. In this study we
have shown that the 10 micron feature is independent of the classification of the
objects. Therefore, we also allowed group I sources in Fig.~\ref{Siemission.ps}.
The correlation plot (Fig.~\ref{FWHMvsPF97.ps}) of the FWHM versus the
$PF/CF$ ratio of the 
amorphous 10 micron silicate feature shows that indeed the FWHM is
larger when the peak-over-continuum flux ratio is lower. Again, no
difference is observed between group I and group II sources. The
dashed line represents the best fit and has a slope of -0.38 micron.

Van Boekel et al. (\citeyear{vanboekel}) suggested the amorphous 10 micron silicate 
emission originates from the disk's surface. It might
appear contradictory that the strength of PAH emission 
\textit{does} depend on the shape of the SED (which represents the disk
geometry), while the 10 micron feature \textit{does
  not}. Nevertheless, the excitation mechanisms for the two
  emission sources are not the same: the non-equilibrium PAH emission
  occurs during temperature fluctuations after absorption of a UV
  photon while the larger silicate grains are in thermal equilibrium
  with the radiation field of the central star.
The models of \citet{habart} show that the PAH emission
emanates mostly from the outer parts of the disk ($\sim$100 \AU),
while the warm silicate emission is confined to the innermost disk
regions ($\sim$ a few \AU). This reflects the different excitation
mechanisms of the small PAH molecules and larger silicate grains, and
is in accordance with the observations: spatially resolved PAH
emission, on scales of $\sim$10--100 \AU, has been detected around HAEBEs
by \citet{geers} and \citet{vanboekel04}.

The results of the present paper
on both the PAH emission and the silicate emission are consistent 
with a scenario in which Group I sources have disks that are 
flaring, whereas Group II sources have flatter disks that are 
shadowed by the disk's puffed-up inner rim.

\section{Conclusions and discussion}

We summarize the conclusions which were drawn in the previous sections.

\begin{enumerate}

\item Although in general the PAH emission-line spectrum of HAEBEs is
fairly homogeneous, we find clear evidence of differences in size,
chemistry and/or ionization for the PAH molecules
within the group of Herbig Ae/Be stars. 

\item The 6.2 micron feature is the most frequently detected PAH feature in our
analysis. However, in 56\% of the spectra where the 6.2
micron feature is detected, the PAH 3.3 micron feature is
\textit{un}detected. Hence the 3.3 micron feature is not a good
indicator for the presence of PAH emission, since quite a large
fraction of the PAH emitters may be overlooked if one only 
considers this feature.

\item PAH emission occurs more frequently, and is stronger in M01
  Group I sources than in Group II sources. If we assume 
that PAH emission originates in the disk's atmosphere, this result 
provides strong support for the interpretation of Group I sources as 
sources with flared disks, and Group II sources as those with 
self-shadowed disks.

\item The strength of the PAH bands is correlated with the
strength of the UV radiation of the central star, but the relative
strength of the PAH emission \textit{decreases} with increasing $L_{\uv}$.
This might be due to the fact that the power absorbed by PAHs
  changes with the spectral energy distribution of the stellar
  radiation field and decreases with increasing hardness.

\item Except for \object{MWC 297}, no PAH emission is observed in group III
sources. 

\item Nanodiamond features were detected in four sample stars, of
which three are group I objects and one is group III.

\item The shape and strength of the 10 micron feature is independent
of the M01 classification; strong, weak and even absent emission
features are observed in both groups. The 10 micron feature in the
spectra of the sample sources indicates that warm amorphous grains are
processed in HAEBEs.

\item Neither the PAH features nor the 10 micron feature
depend on disk mass. The masses of the disks are also independent of the
classification.

\item The 11 micron feature does not correlate with the amorphous
silicate feature, which could indicate that the PAHs are the dominant
ingredient of the feature and/or that crystalline silicates have an
independent evolution. The 11 micron complex is in general stronger in
group I sources. 

\item Crystalline silicates in emission occur in both types of
sources. \object{Z CMa} is the only source where the feature is possibly seen
in \textit{absorption}.

\end{enumerate}

We confirm in the present paper that the strongest PAH-emission occurs
in group I sources, as was first noted by M01. This is 
consistent with the idea that group I sources have a flared disk
geometry, while group II sources are systems with self-shadowed disks 
\citep{dullemond02}. In a flared disk geometry, the UV photons of the
central star can reach the PAH 
molecules in the surface layers of the disk in a direct way. PAH
molecules in the atmosphere of self-shadowed disks are shaded from the UV field by
the puffed-up inner rim which casts its shadow over the outer parts of
the disk. Therefore, the PAH spectrum of group II sources is expected
to be much fainter than that of group I sources. Most sample stars follow
this hypothesis. 

The amorphous 10 micron silicate feature in this sample of HAEBE stars
displays the characteristics of grain growth, independent of the
classification of the sources. The small hot silicate grains from
which this feature originates seem to have little interaction with
the PAH molecules, even though they are believed to radiate from the
same locus: the disk atmosphere. Crystallinity seems to be present
in more than 1/4 of the stars in our sample. 

The global interpretation of the SED of HAEBEs in terms of disk
geometry is in many ways consistent with the observations. Future work
with new generation IR instruments (e.g. Spitzer) and interferometers
(e.g. MIDI/AMBER on the VLTI) will 
provide the community with new high-quality observations of HAEBE stars,
which will offer a good test for the hypotheses suggested in this
paper.

\section*{Appendix}

We estimate the density of PAH molecules in a spherical, optically
thin halo of radius $R$ around the central star, supposing a homogeneous
distribution of the PAHs. Furthermore, we assume that all UV radiation
absorbed by the PAH molecules is re-emitted in the IR bands.

The cross-section of a single PAH molecule with 50 C atoms is
typically $\sigma_C = 2.5 \times 10^{-14}~\cm^2$ \citep{li02}.
The luminosity of the PAH emission $L_{\mathrm{PAH}}$
depends on the number of emitting PAHs $N_{\mathrm{PAH}}$, the average
energy of the absorped UV photons $\langle E_{\uv} \rangle$, the
number of UV photons $N_{\uv}$ that pass by per second and 
the cross-section of a single PAH molecule over the average surface $S$ of
the sphere:
\begin{eqnarray}  
L_{\mathrm{PAH}} & = & N_{\mathrm{PAH}} \langle E_{\uv} \rangle N_{\uv} \frac{\sigma_C}{S}      \nonumber \\
        & = & \frac{4\pi}{3} R^3 \rho \times \langle E_{\uv} \rangle \times\
        \frac{L_{\uv}}{\langle E_{\uv} \rangle} \times \frac{\sigma_C}{S}    \nonumber \\
\frac{L_{\mathrm{PAH}}}{L_{\uv}}        & = & 0.37\
        \rho[\cm^{-3}]\ R[\AU]     \label{lpaht}
\end{eqnarray}
Assuming that all IR PAH emission is observed in the measured PAH bands,
the sum of the line fluxes of these features is connected to the PAH luminosity
by
\begin{eqnarray}  
L_{\mathrm{PAH}} [L_{\odot}] & = & 2.4\ 10^{6} \sum_{\mathrm{PAH~j}} 4\pi (d[pc])^2
\times LF_j \biggl[ \frac{W}{m^2} \biggr]  \label{lpaho}  \\
                    & = & \sum_{\mathrm{PAH~j}} L_{\mathrm{PAH~j}} [L_{\odot}] \label{lpahi}
\end{eqnarray}
in which the terms of the sum are the luminosities of the individual
PAH features.
Hence from formulas~\ref{lpaht} and \ref{lpaho}, an estimate for
$\rho$ can be derived, assuming that the radius of the halo $R$ is
equal to the radius of the ISO-SWS beam (11\arcsec) at the distance $d$ of the
source, or
\[ R[\AU] = 11\ d[pc]   \]

A typical value for the ISM abundance of PAHs over H atoms
($[\mathrm{PAH}/\mathrm{H}]=10^{-7}$) can be found in
\citet{tielens99}. Typical ISM electron densities $\rho_{e^-}$ are in
the order of 5 cm$^{-3}$. Hence a typical value for the ISM density of
PAH molecules is 
\begin{eqnarray}
  \rho_{\mathrm{typ}} = \biggl[\frac{\mathrm{PAH}}{\mathrm{H}}\biggr] \times \rho_{e^-} = 5 \times
  10^{-7} \cm^{-3}
\end{eqnarray}

\acknowledgements{
The authors would like to thank Drs. C. Dullemond, R. van Boekel 
and R. Waters for many useful discussions that led to significant 
improvements in the manuscript.  We are also indebted to 
Dr. B. Vandenbussche for his help with the reduction of the 
ISO-SWS spectra.
This publication makes use of data products from the Two Micron All Sky 
Survey, which is a joint project of the University of Massachusetts and 
the Infrared Processing and Analysis Center, funded by the National 
Aeronautics and Space Administration and the National Science Foundation. 
This research has 
made use of the Simbad data base, operated at CDS, Strasbourg, France.}

\bibliographystyle{aa}
\bibliography{references.bib}

\end{document}